\documentclass[useAMS,usenatbib]{mn2e}

\usepackage{rotating}
\usepackage{longtable}
\usepackage{graphics}
\usepackage{aas_macros}
\usepackage{booktabs,caption}

\newcommand{\teff}{$T_{\rm eff}$}

\newcommand{\eexc}{$E_{\rm exc}$}
\def\lgg{log\,${g}$}
\def\vt{$\xi_{\rm t}$}

\newcommand{\kms}{km\,s$^{-1}$}

\def\ione{\,{\sc i}}
\def\ii{\,{\sc ii}}
\def\iii{\,{\sc iii}}
\newcommand{\eps}{\log\varepsilon}

\newcommand*\rfrac[2]{{}^{#1}\!/_{#2}}
\newcommand{\hi}{H\,{\textsc i}\,}
\newcommand{\zni}{Zn\,{\textsc i}\,}

\title[]{Non-LTE abundances of zinc in different spectral type stars and the Galactic [Zn/Fe] trend based on quantum-mechanical data on inelastic processes in zinc-hydrogen collisions}
\author[T. M. Sitnova et al.]{T. M. Sitnova$^{1}$\thanks{E-mail:sitnova@inasan.ru}; S. A. Yakovleva$^{2}$, A. K. Belyaev$^{2}$, L. I. Mashonkina$^{1}$ \\
$^{1}$Institute of Astronomy, Russian Academy of Sciences, Pyatnitskaya 48, 119017, Moscow, Russia\\
$^{2}$Department of Theoretical Physics and Astronomy, Herzen University, St. Petersburg, 191186, Russia \\
}

\begin{document}

\date{}

\pagerange{\pageref{firstpage}--\pageref{lastpage}} \pubyear{2021}

\maketitle

\label{firstpage}

\begin{abstract}

We present a new model atom of Zn\ione -Zn\ii\ based on the most up-to-date photoionisation cross-sections, electron-impact excitation
rates, and rate coefficients for the Zn\ione\ + H\ione\ and Zn\ii\ + H$^-$ collisions. The latter were calculated using the multi-channel quantum asymptotic treatment based on the Born-Oppenheimer approach. Non-LTE analysis was performed for the first time for lines of Zn\ione\ and Zn\ii\ in the ultraviolet (UV) spectra of two very metal-poor reference stars, HD~84937 and HD~140283. We found consistent non-LTE abundance from the resonance Zn\ione\ 2138 \AA\ line, the subordinate lines, and the lines of Zn\ii. In both stars, non-LTE leads to 0.17~dex higher average abundance from Zn\ione, while, for Zn\ii\ lines, non-LTE corrections are minor and do not exceed 0.06~dex. Using lines of Zn\ione\ in the high-resolution spectra, we determined the non-LTE abundances for a sample of 80 stars in the $-2.5 \le$ [Fe/H] $\le$ 0.2 metallicity range. The [Zn/Fe] versus [Fe/H] diagram reveals a dip, with [Zn/Fe] $\simeq$ 0.3 in the most metal-poor stars, a close-to-solar value for [Fe/H] $\sim -1.2$, and increasing [Zn/Fe] up to 0.3 in the thick disk stars. The close-to-solar metallicity stars have subsolar  [Zn/H] $\simeq -0.1$, on average. Non-LTE abundances of zinc were derived for the first time for seven reference F to B-type stars. We provide a grid of the non-LTE abundance corrections.

\end{abstract}


\section{Introduction}

Along with iron, zinc is a pivotal chemical element in astrophysics. It is treated as  an iron peak element, however, in contrast to iron,  it is volatile and it does not settle in dust grains \citep{Lodders2009}. For this reason, zinc has become a tracer of metallicity in studies of interstellar medium, namely, distant protogalaxies at the early stage of chemical evolution known as  damped Ly$\alpha$ systems \citep[DLAs, see, for example,][]{2005ARA&A..43..861W,2011A&A...530A..33V,2016MNRAS.463.3021B}. 
Studies of gas and dust chemical properties  in DLAs with different metallicity invoke the observed Galactic [Zn/Fe]\footnote{We use a standard designation, [X/Y] = log(N$_{\rm X}$/N$_{\rm Y}$)$_{*}$ - log(N$_{\rm X}$/N$_{\rm Y}$)$_{\odot}$, where N$_{\rm X}$ and N$_{\rm Y}$ are total number densities of element X and Y, respectively.} -- [Fe/H] trend.
Thus, an accuracy of zinc  abundance  determination in low mass long-lived stars with different [Fe/H] impacts understanding chemical evolution either  in the Milky Way as well as in  other distant galaxies. 

A lot of studies have been performed to establish [Zn/Fe] -- [Fe/H] Galactic trend \citep{1988ApJ...335..406S,1991A&A...246..354S,2000LIACo..35..119P,2002A&A...396..189M,2003MNRAS.340..304R,2003A&A...410..527B,2004A&A...415..993N,2004A&A...425..697C,2004A&A...416.1117C,2005A&A...433..185B,Takeda2005zn,2009PASJ...61..549S,
2014A&A...562A..71B,
2017A&A...604A.128D,2017A&A...600A..22M,2018ApJ...857....2R}. 
The majority of the above studies report on a moderate overabundance  in [Zn/Fe] in stars with [Fe/H]$< -1$ and decreasing [Zn/Fe] at higher metallicity.   Some studies managed to detect a separation and higher [Zn/Fe] in thick disk stars compared to thin disk stars with similar [Fe/H] \citep{2003A&A...410..527B,2017A&A...600A..22M}. Such a behaviour is qualitatively similar to those observed for  $\alpha$-elements. Other details in the zinc trend were figured out: \citet{2002A&A...396..189M}, \citet{2004A&A...415..993N}, and \citet{2009PASJ...61..549S}  found a pit in [Zn/Fe] at [Fe/H] $\simeq -1.2$. At solar [Fe/H], \citet{2017A&A...604A.128D} found lower [Zn/Fe] ratios in stars at Galactocentric distances less than 7.5 kpc. What for very metal-poor (VMP, [Fe/H] $<-2$) tail of the trend, [Zn/Fe] increases with decreasing metallicity [Fe/H] $< -2.5$ and can reach [Zn/Fe] = 1~dex at [Fe/H] = $-4$ \citep{2000LIACo..35..119P,2004A&A...416.1117C,Takeda2005zn}. 

Zinc abundance determination in the above studies rely on the assumption of local thermodynamic equilibrium (LTE). An exception is \citet[][hereafter, T05]{Takeda2005zn} and \citet{2016PASJ...68...81T}, who studied an impact of non-LTE effects on  Zn\ione\ 4722, 4810, and 6362 \AA\ lines in a wide metallicity range of late type stars. T05 found non-LTE abundance corrections ($\Delta_{\rm NLTE}$, the difference between non-LTE and LTE abundances for individual spectral lines) to be moderate in absolute value, within 0.1~dex. However, the value and sign of non-LTE corrections depend on stellar parameters and metallicity. For example, for the Zn\ione\ 4810 \AA\ line in the Sun, T05 found   $\Delta_{\rm NLTE} = -0.05$, while, for metal-poor stars, non-LTE abundance corrections are mainly positive. Thus, neglecting the non-LTE effects may lead to a shift in [Zn/H] ratio of 0.15~dex and distort [Zn/Fe] -- [Fe/H] abundance trend.

Not only accurate line formation, but also an accurate homogeneous set of stellar atmosphere parameters (effective temperature, surface gravity, and metallicity) are required to establish a robust [El/Fe] trend. T05 compiled equivalent width measurements and stellar atmosphere parameters from different studies, determined by different methods. For example, \citet{2004A&A...416.1117C} determined surface gravities for a sample of metal-poor stars using Fe\ione\--Fe\ii\ and Ti\ione\--Ti\ii\ ionisation balance in LTE, while \citet{2004A&A...415..993N} adopted distance based surface gravities. While, in solar type dwarfs, non-LTE weakly affect Fe\ione\--Fe\ii\ and Ti\ione\--Ti\ii\ abundance difference, in metal-poor stars, non-LTE effects are significant for lines of neutral species, and an application of ionisation balance method based on LTE analysis may lead to uncertainty in surface gravity of 0.5~dex in dwarfs \citep[see, for example,][]{lick} and to even larger effects in giants \citep[see, for example,][]{mash_fe}. Adopting stellar parameters determined by different methods can produce a systematic offset between abundances from different stellar subsamples. 

At metallicity [Fe/H] $<-3$, Zn\ione\ lines in the visible spectrum range (4680, 4722, 4810, and 6362 \AA) became weaker and hardly can be detected. To determine zinc abundance in stars with extremely low metallicities, one have to invoke the Zn\ione\ and Zn\ii\ resonance line observed in the ultraviolet spectrum (UV) range \citep{2016ApJ...824L..19R,2018ApJ...857....2R,2019ApJ...876...97E}. For six metal-poor stars ($-2.5 <$ [Fe/H]$< -1$), \citet{2018ApJ...857....2R} found a good agreement between the average abundance from Zn\ione\ lines in the visible and UV range and the resonance Zn\ii\ 2062 \AA\ line. \citet{2018ApJ...857....2R} reported significant abundance uncertainties of 0.2~dex for the UV lines, and, in metal-poor reference stars with well-determined stellar parameters, abundances between different Zn\ione\ lines differ up to 0.22~dex. 
In order to safely adopt these lines for abundance determinations in extremely MP stars, accurate line formation calculations should be performed.

Lines of Zn\ione\ are observed also in the visible spectrum range in A and late B-type stars. 
With the ionisation energy E$_{ion}$(Zn\ione) = 9.39~eV, in AB-type stars, Zn\ione\ is a minority species and one can expect significant non-LTE effects. 
To our knowledge, the non-LTE calculations have not been performed yet for zinc in hot stars.
To estimate an impact of non-LTE on abundance determination detailed calculations are required.

We constructed the Zn\ione--\ii\ model atom using the most accurate atomic data for electronic collisions, collisions with hydrogen atoms, and photoionisation cross-section. The method of non-LTE abundance determination, atomic data, and codes are described in Sect.~\ref{method}. The mechanism of the departures from LTE are described in Sect.~\ref{se}. We  tested the non-LTE method for zinc by analysis of different Zn\ione\ and Zn\ii\ lines in reference stars (Sect.~\ref{testing}). 
Non-LTE abundances of zinc were derived for seven F to late B-type stars (Sect.~\ref{baf}).
Using a sample of dwarf stars in a wide metallicity range and robust stellar parameters, we revisited the Galactic [Zn/Fe] -- [Fe/H] trend (Sect.~\ref{znfe}). To account for the departures from LTE, we provide non-LTE abundance corrections for different zinc lines (Sect.~\ref{grid}). Our results  are summarised in Sect.~\ref{conclusions}.

\section{Abundance determination method}\label{method}

\subsection{Model atom}\label{atom}

\begin{figure}
	\includegraphics[width=80mm]{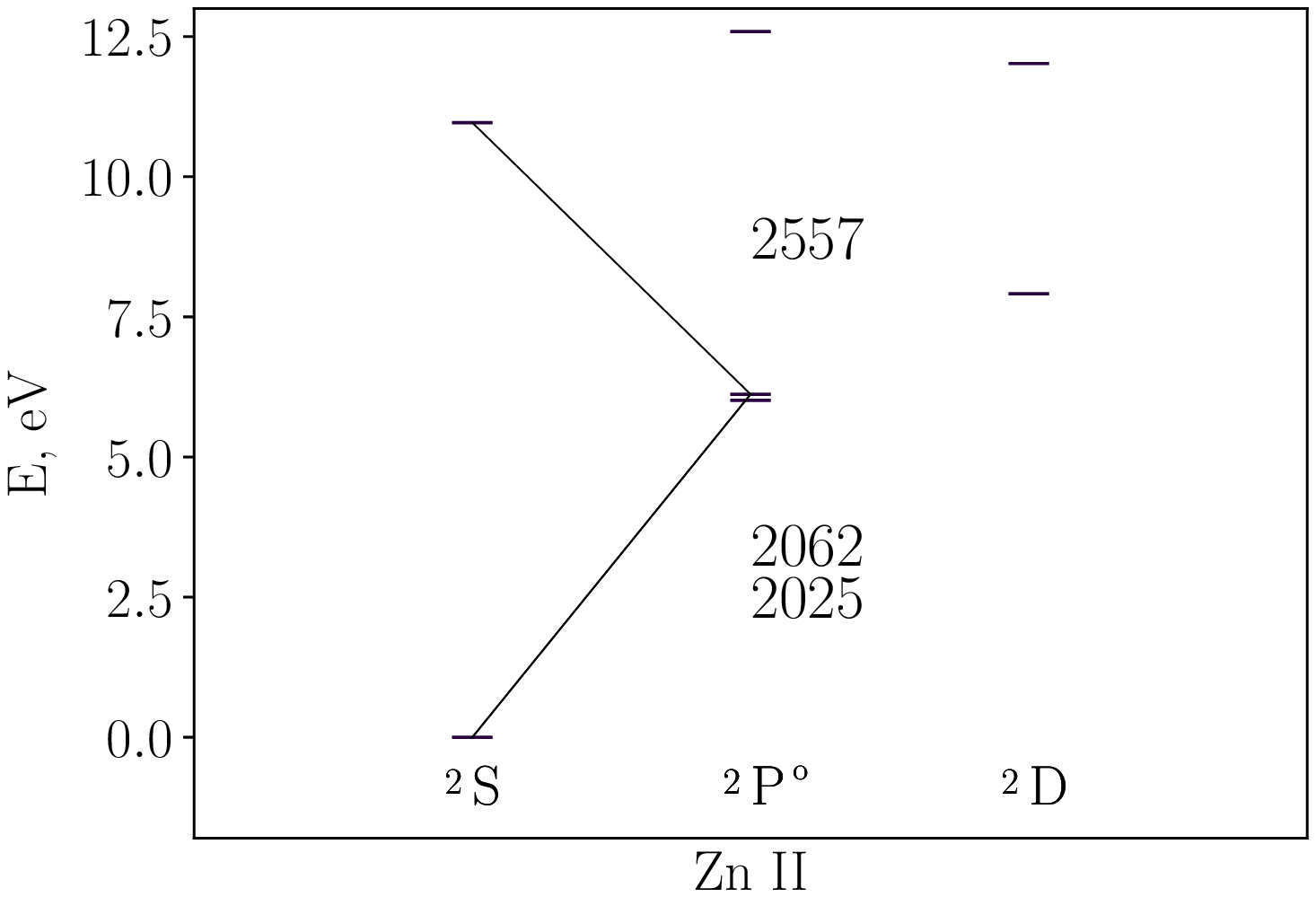}
	\includegraphics[width=80mm]{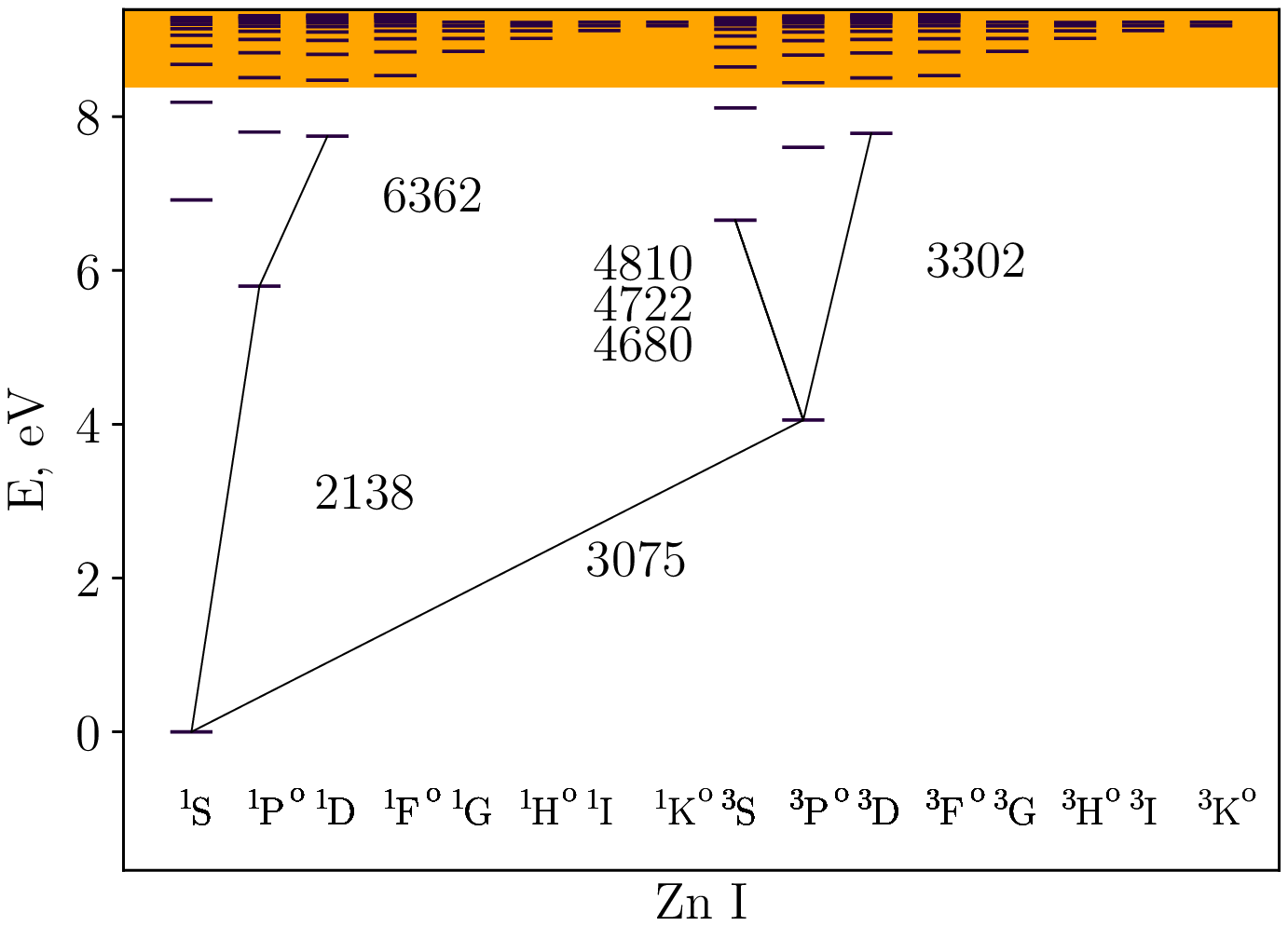}
	\caption{	
Zn\ione\ (bottom panel) and Zn\ii\ (top panel) grotrian diagrams. The high-excitation levels of Zn\ione\ marked with shaded area are combined into the seventeen superlevels. The wavelengths of the observed spectral lines are indicated in \AA.}
	\label{zn12atom} 
\end{figure}

\underline{Energy levels.}
Our Zn\ione--\ii\ model atom (Fig.~\ref{zn12atom}) includes 28 levels of Zn\ione, 7 levels of Zn\ii, and the ground state of Zn\iii. The list of energy levels and transitions is taken from R.~Kurucz webpage\footnote{http://kurucz.harvard.edu/atoms.html}.
{For the atomic levels with \eexc\ up to 9~eV their energies are taken from the laboratory measurements and implemented in the atomic structure calculations by R.~Kurucz.}
For Zn\ione, we included into the model atom all levels up to the ionization threshold, 9.39~eV. Levels with an excitation energy larger than 8.2~eV are combined into the superlevels with energy separation of 0.1~eV. Fine structure is neglected for Zn\ione. 
Zn\ii\ model atom includes seven levels with the excitation energy up to 12.6~eV. For 4p~$^2$P$^{\circ}$ level, we take into account fine structure, since the splitting is significant and amounts to 0.11~eV.

\underline{Radiative transitions.}
The atomic levels in the model atom are connected by radiative and collisional bound-bound and bound-free transitions. In the statistical equilibrium (SE) calculations, the oscillator strength are taken from R.~Kurucz database. 
For strong transitions that correspond to the observed in stellar spectra lines, we adopted in SE calculations the oscillator strengths from laboratory measurements of \citet{1975JQSRT..15...25A} and \citet{2012ApJ...750...76R} for Zn\ione, and from \citet{2006EPJD...37..181M} for Zn\ii. 
For accurate calculation of radiative rates in the resonance transitions of Zn\ione\ and Zn\ii, we use the Voigt profile and a fine wavelength mesh, which fully covers the lines core and wings.

\begin{figure}
	\includegraphics[width=80mm]{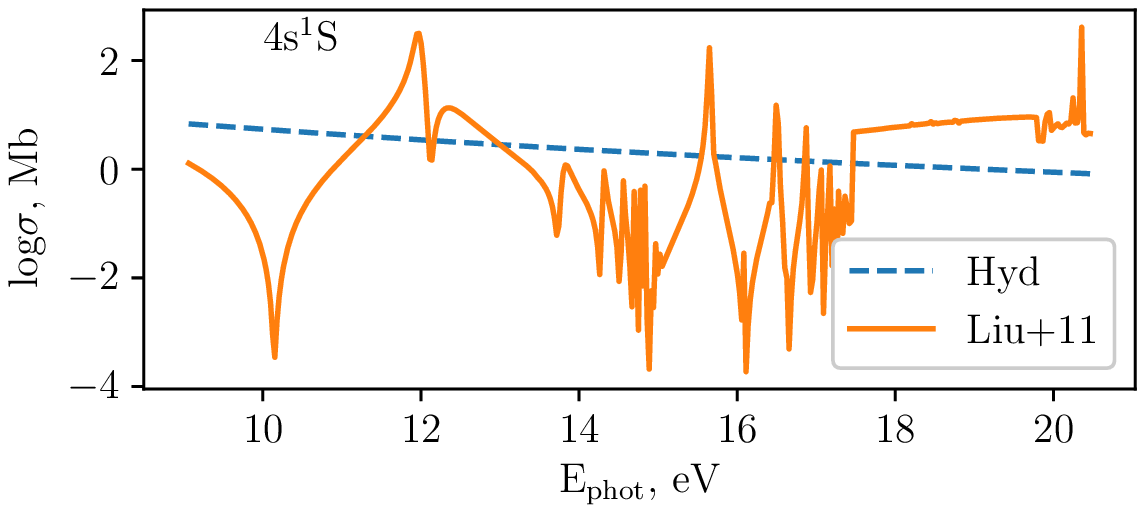}
	\includegraphics[width=80mm]{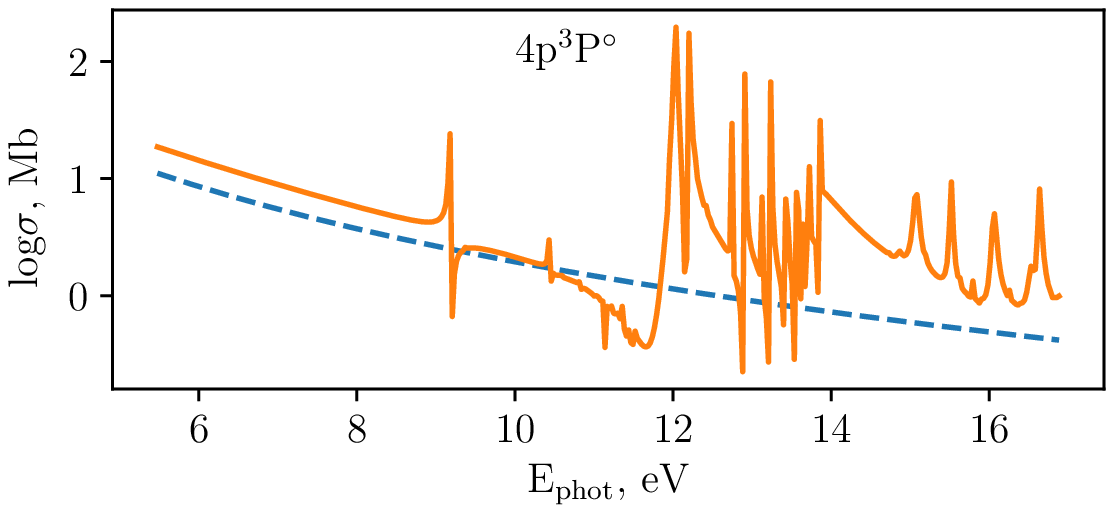}
	\includegraphics[width=80mm]{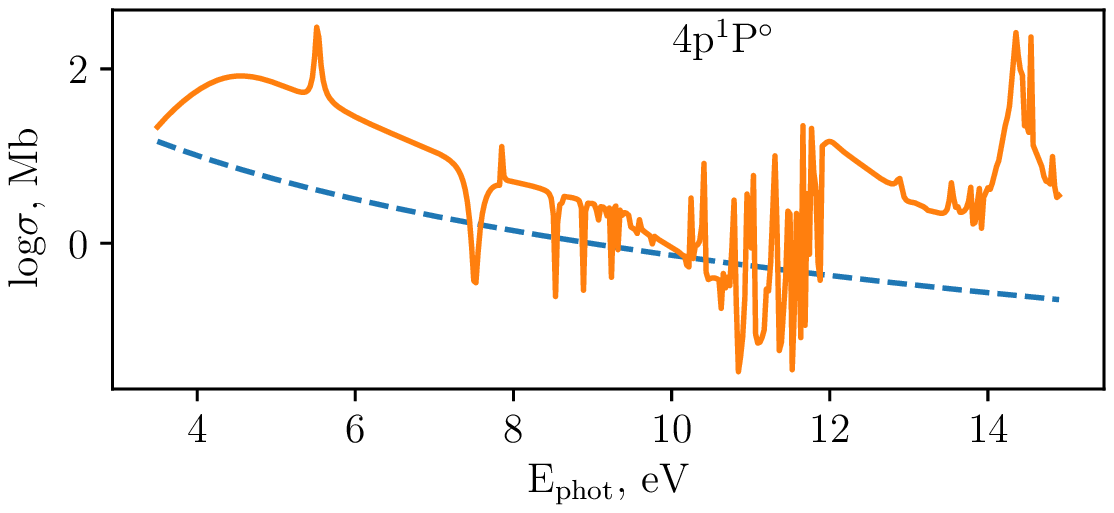}	
	\includegraphics[width=80mm]{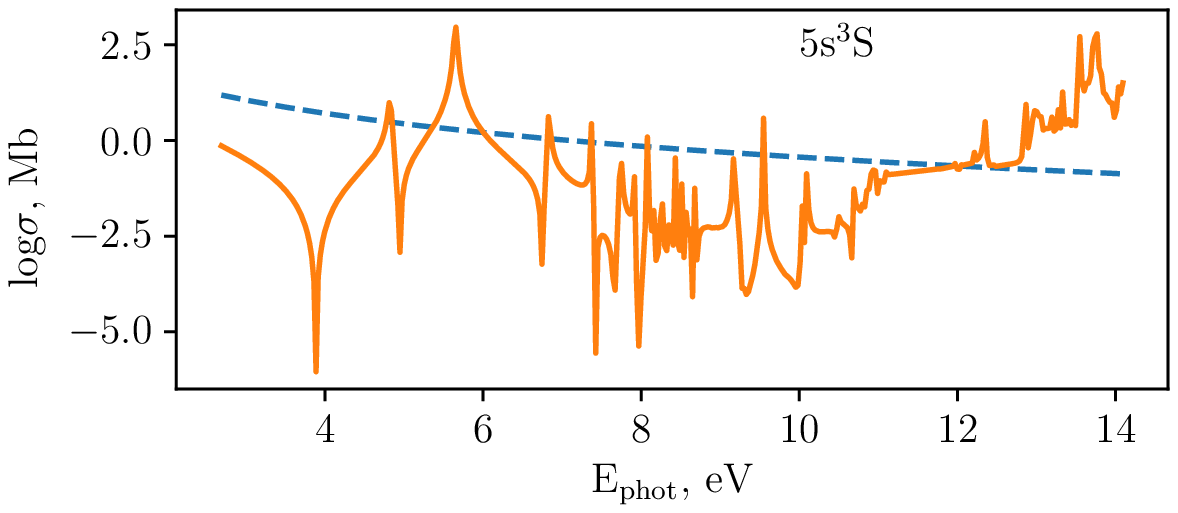}	
	\includegraphics[width=80mm]{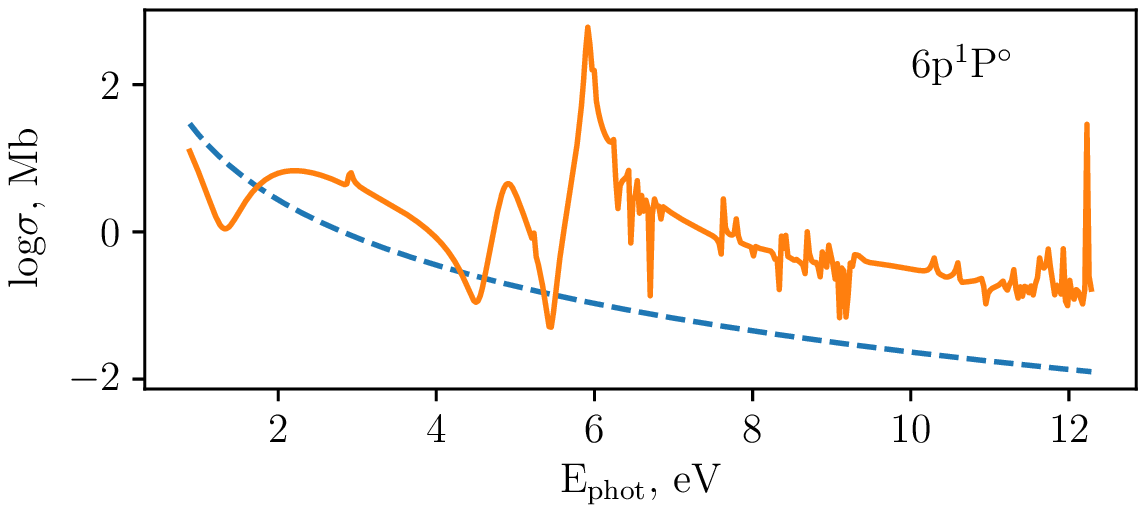}	
	\caption{Photoionisation cross-sections for the selected levels of Zn\ione\ from quantum mechanical calculations of \citet[][solid lines]{2011A&A...536A..51L} and computed in the hydrogenic approximation (dashed lines).}
	\label{pic_liu} 
\end{figure}

For Zn\ione\ terms with \eexc $\le$ 8.2~eV, we use photoionisation cross-sections from the R-matrix calculations of \citet{2011A&A...536A..51L}. For the remaining high-excitation levels, we assume a hydrogenic approximation with using an effective principle quantum number. We compare the quantum mechanical photoionisation cross-sections with the hydrogenic ones for selected levels Zn\ione\ in Fig.~\ref{pic_liu}. 

For all levels of Zn\ii, we calculated photoionization cross-sections using hydrogenic approximation with the effective principle quantum number. Zn\ii\ is a majority species in the stellar parameter range with which we concern. Therefore, the ionisation / recombination processes nearly do not affect the SE of Zn\ii. 

\begin{figure}
	\includegraphics[width=80mm]{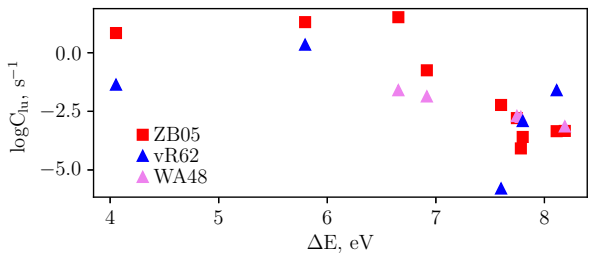}
	\caption{	
		Electron-impact excitation rates for the Zn\ione\ transitions at T = 5000 K and N$_e$ = 1.05E+12 (these values correspond to log$\tau_{5000} = -1$ in model atmosphere with \teff/ log~g/ [Fe/H] = 5780/3.70/--2.46) calculated from \citet{2005PhRvA..71b2716Z} data (squares) and formulae of \citet{Reg1962} and \citet{1948MNRAS.108..292W} (blue and violet triangles, respectively).  }
	\label{ar_z} 
\end{figure}

\underline{Collisions with electrons.}
For electronic collisions of Zn\ione, we used the cross sections generated in the R-matrix calculations described by \citet{2005PhRvA..71b2716Z}. 
These calculations were performed for ten transitions from the ground state of Zn\ione\ to the levels up to 6s$^1$S with \eexc = 8.2~eV.
The rate coefficients were computed by integrating the cross-sections over the Maxwellian velocity distribution for electrons, and they are presented in Table~\ref{e_rates}. 
For the ten transitions, Fig.~\ref{ar_z} shows a  comparison of electron-impact excitation rates derived from \citet{2005PhRvA..71b2716Z} data and those calculated with the approximate \citet{Reg1962} and \citet{1948MNRAS.108..292W} formulae, for radiatively allowed and forbidden transitions, respectively. 
The above approximate formulae were adopted for electron collision rates calculations for the remaining  Zn\ione\ and Zn\ii\ transitions.
Electron impact ionisation is calculated with the \citet{1962amp..conf..375S} formula.

\begin{table*}
	\begin{center}
		\caption{Electron-impact excitation rate coefficients (in cm$^3$ s$^{-1}$) for the Zn\ione\ transitions at the
		temperatures from 2000 K to 50000 K with a step of 2000 K.}
		\label{e_rates} 
		\setlength{\tabcolsep}{0.7mm}
		\begin{tabular}{l}
			\hline
4s $^1$S --  4p $^3$P$^{\circ}$  \\
0.7713E-17 0.9732E-12 0.4484E-10 0.2891E-09 0.8547E-09 0.1718E-08 0.2776E-08 0.3922E-08 0.5071E-08 0.6167E-08 \\
0.7178E-08 0.8088E-08 0.8892E-08 0.9591E-08 0.1019E-07 0.1070E-07 0.1113E-07 0.1148E-07 0.1177E-07 0.1199E-07 \\
0.1217E-07 0.1230E-07 0.1239E-07 0.1245E-07 0.1247E-07 \\
4s $^1$S --  4p $^1$P$^{\circ}$  \\
0.2192E-16 0.2832E-11 0.1316E-09 0.8526E-09 0.2528E-08 0.5091E-08 0.8239E-08 0.1165E-07 0.1508E-07 0.1835E-07 \\
0.2137E-07 0.2409E-07 0.2650E-07 0.2859E-07 0.3040E-07 0.3193E-07 0.3321E-07 0.3427E-07 0.3513E-07 0.3581E-07 \\
0.3634E-07 0.3674E-07 0.3701E-07 0.3718E-07 0.3727E-07 \\
4s $^1$S --  5s $^3$S \\  
0.3423E-16 0.4570E-11 0.2147E-09 0.1398E-08 0.4157E-08 0.8388E-08 0.1359E-07 0.1925E-07 0.2493E-07 0.3036E-07 \\
0.3537E-07 0.3990E-07 0.4390E-07 0.4739E-07 0.5039E-07 0.5294E-07 0.5507E-07 0.5684E-07 0.5828E-07 0.5942E-07 \\
0.6031E-07 0.6097E-07 0.6144E-07 0.6173E-07 0.6188E-07 \\
4s $^1$S --  5s $^1$S \\  
0.5053E-21 0.9998E-14 0.2888E-11 0.5077E-10 0.2895E-09 0.9379E-09 0.2194E-08 0.4181E-08 0.6934E-08 0.1042E-07 \\
0.1454E-07 0.1919E-07 0.2424E-07 0.2955E-07 0.3502E-07 0.4053E-07 0.4600E-07 0.5136E-07 0.5654E-07 0.6151E-07 \\
0.6624E-07 0.7070E-07 0.7489E-07 0.7879E-07 0.8240E-07 \\
4s $^1$S --  5p $^3$P$^{\circ}$  \\
0.9700E-24 0.2377E-15 0.1340E-12 0.2994E-11 0.1858E-10 0.6122E-10 0.1409E-09 0.2599E-09 0.4141E-09 0.5961E-09 \\
0.7975E-09 0.1010E-08 0.1227E-08 0.1443E-08 0.1654E-08 0.1856E-08 0.2048E-08 0.2227E-08 0.2394E-08 0.2548E-08 \\
0.2688E-08 0.2816E-08 0.2931E-08 0.3034E-08 0.3127E-08 \\
4s $^1$S --  4d $^1$D \\  
0.9185E-25 0.5545E-16 0.4341E-13 0.1178E-11 0.8464E-11 0.3149E-10 0.8066E-10 0.1637E-09 0.2841E-09 0.4419E-09 \\
0.6338E-09 0.8549E-09 0.1099E-08 0.1360E-08 0.1632E-08 0.1910E-08 0.2189E-08 0.2464E-08 0.2733E-08 0.2994E-08 \\
0.3244E-08 0.3481E-08 0.3706E-08 0.3917E-08 0.4115E-08 \\
4s $^1$S --  4d $^3$D \\  
0.9982E-27 0.2279E-17 0.2746E-14 0.9187E-13 0.7412E-12 0.2950E-11 0.7853E-11 0.1627E-10 0.2853E-10 0.4451E-10 \\
0.6378E-10 0.8574E-10 0.1097E-09 0.1351E-09 0.1612E-09 0.1876E-09 0.2138E-09 0.2396E-09 0.2645E-09 0.2884E-09 \\
0.3112E-09 0.3327E-09 0.3529E-09 0.3718E-09 0.3894E-09 \\
4s $^1$S --  5p $^1$P$^{\circ}$  \\
0.3046E-26 0.7014E-17 0.8458E-14 0.2826E-12 0.2276E-11 0.9041E-11 0.2403E-10 0.4971E-10 0.8706E-10 0.1357E-09 \\
0.1942E-09 0.2609E-09 0.3336E-09 0.4105E-09 0.4897E-09 0.5696E-09 0.6489E-09 0.7267E-09 0.8020E-09 0.8742E-09 \\
0.9430E-09 0.1008E-08 0.1069E-08 0.1126E-08 0.1179E-08 \\
4s $^1$S --  6s $^3$S \\  
0.5336E-26 0.1240E-16 0.1492E-13 0.4965E-12 0.3979E-11 0.1574E-10 0.4165E-10 0.8587E-10 0.1499E-09 0.2330E-09 \\
0.3328E-09 0.4461E-09 0.5695E-09 0.6997E-09 0.8335E-09 0.9684E-09 0.1102E-08 0.1233E-08 0.1360E-08 0.1481E-08 \\
0.1597E-08 0.1706E-08 0.1808E-08 0.1904E-08 0.1992E-08 \\
4s $^1$S --  6s $^1$S \\  
0.2987E-26 0.1139E-16 0.1704E-13 0.6606E-12 0.6011E-11 0.2656E-10 0.7759E-10 0.1746E-09 0.3294E-09 0.5485E-09 \\
0.8326E-09 0.1178E-08 0.1577E-08 0.2022E-08 0.2502E-08 0.3006E-08 0.3527E-08 0.4055E-08 0.4583E-08 0.5105E-08 \\
0.5615E-08 0.6110E-08 0.6587E-08 0.7043E-08 0.7477E-08 \\
			\hline
\end{tabular}
\end{center}
\end{table*}

\underline{Collisions with hydrogen atoms.}

For SE calculations of \zni\ in late-type stars, inelastic processes in collisions with \hi\ atoms are taken into account. We have calculated for the first time the rate coefficients for bound-bound transitions in inelastic collisions of zinc atoms and positive ions with hydrogen atoms and negative ions.
The calculations are performed using the multichannel quantum asymptotic treatment proposed by \cite{Belyaev-pra-2013} within the Born-Oppenheimer approach. 

The present investigation of inelastic processes of excitation, de-excitation, mutual neutralization and ion pair formation includes 12 scattering channels: 11 channels asymptotically correspond to covalent molecular states ${\rm Zn}(3d^{10}4s\,nl~^{1,3}\!L) + {\rm H}(1s~^2S)$ and one to ionic molecular state ${\rm Zn}^{+}(3d^{10}4s~^2S) + {\rm H}^-(1s^2~^1S)$. The ionic molecular state forms $^2\Sigma^+$ molecular symmetry, and only those covalent states that have $^2\Sigma^+$ symmetry are included into consideration. All the states are collected in Table~\ref{tab:states} together with their asymptotic energies and statistical probabilities  for population of $^2\Sigma^+$ molecular states.
For the considered molecular states the diabatic Hamiltonian matrix is constructed, diagonalization of this matrix gives the adiabatic molecular potentials.

\begin{table}
	\begin{center}
		\caption{ZnH $(k~^2\Sigma^+)$ molecular states (in the $LS$ representation), the corresponding scattering channels, their asymptotic energies with respect to the ground-state from NIST \citep{NIST}, and the statistical probabilities $p_k^{stat}$ for population of the molecular states $^2\Sigma^+$.}
		\label{tab:states} 
	\setlength{\tabcolsep}{0.7mm}
\renewcommand{\arraystretch}{1.8} 
		\begin{tabular}{lllc}
			\hline
			k &  Scattering channels  & Asymptotic    & $p_k^{stat}$ \\[-2mm] 
			&                                      & energies (eV) &   \\
			\hline
			1  & ${\rm Zn}(3d^{10}4s^{2}~^1S) + {\rm H}(1s~^2S)$ & 0.000000 & 1 \\ 
			2  & ${\rm Zn}(3d^{10}4s4p~^3P^{\circ}) + {\rm H}(1s~^2S)$ &  4.077881 & $\rfrac{1}{9}$ \\
			3  & ${\rm Zn}(3d^{10}4s4p~^1P^{\circ}) + {\rm H}(1s~^2S)$ &  5.795691 & $\rfrac{1}{3}$ \\
			4  & ${\rm Zn}(3d^{10}4s5s~^3S) + {\rm H}(1s~^2S)$ &  6.654509 & $\rfrac{1}{3}$ \\
			5  & ${\rm Zn}(3d^{10}4s5s~^1S) + {\rm H}(1s~^2S)$ &   6.916981  & 1 \\
			6  & ${\rm Zn}(3d^{10}4s5p~^3P^{\circ}) + {\rm H}(1s~^2S)$ & 7.604056 & $\rfrac{1}{9}$ \\
			7 & ${\rm Zn}(3d^{10}4s4d~^1D) + {\rm H}(1s~^2S)$ & 7.743871 & $\rfrac{1}{5}$ \\ 
			8 & ${\rm Zn}(3d^{10}4s4d~^3D) + {\rm H}(1s~^2S)$ & 7.783354 & $\rfrac{1}{15}$ \\
			9 & ${\rm Zn}(3d^{10}4s5p~^1P^{\circ}) + {\rm H}(1s~^2S)$ & 7.799899 & $\rfrac{1}{3}$ \\
			10 & ${\rm Zn}(3d^{10}4s6s~^3S) + {\rm H}(1s~^2S)$ & 8.112569 & $\rfrac{1}{3}$ \\
			11 & ${\rm Zn}(3d^{10}4s6s~^1S) + {\rm H}(1s~^2S)$ & 8.187627 & 1 \\
			12 & ${\rm Zn}^{+}(3d^{10}4s~^2S) + {\rm H}^-(1s^2~^1S)$ & 8.640197 & 1 \\
			\hline
		\end{tabular}
	\end{center}
\end{table}

\begin{table*}
	\begin{center}
		\caption{Rate coefficients (in cm$^3$ s$^{-1}$) for excitation, de-excitation, neutralization and ion-pair formation processes in ${\rm Zn} + {\rm H}$ and ${\rm Zn}^+ + {\rm H}^-$ collisions at different temperatures.}
		\label{tab:rates} 
		\setlength{\tabcolsep}{0.6mm}
		\renewcommand{\arraystretch}{1.3} 
		\begin{tabular}{lllllllllllll}
			\hline
init. $\downarrow$ fin. $\rightarrow$ & 4s $^1$S & 4p $^3$P$^{\circ}$ & 4p $^1$P$^{\circ}$ & 5s $^3$S & 5s $^1$S & 5p $^3$P$^{\circ}$ & 4d $^1$D & 4d $^3$D & 5p $^1$P$^{\circ}$ & 6s $^3$S & 6s $^1$S & Zn$^+$ + H$^-$\\
		\hline
 \multicolumn{13}{l}{T = 1000 K} \\
4s $^1$S & -- & 7.48E-38 & 7.82E-49 & 6.58E-54 & 4.36E-56 & 7.46E-59 & 9.85E-60 & 5.00E-60 & 3.81E-60 & 6.95E-62 & 3.35E-62 & 2.12E-62\\
4p $^3$P$^{\circ}$ & 2.96E-18 & -- & 1.29E-21 & 2.94E-27 & 1.25E-29 & 1.25E-32 & 1.53E-33 & 7.61E-34 & 5.70E-34 & 1.67E-35 & 9.57E-36 & 1.89E-36\\
4p $^1$P$^{\circ}$ & 4.22E-20 & 1.76E-12 & -- & 8.04E-15 & 2.12E-17 & 1.16E-20 & 1.32E-21 & 6.40E-22 & 4.74E-22 & 1.62E-23 & 1.06E-23 & 1.13E-24\\
5s $^3$S & 7.55E-21 & 8.54E-14 & 1.71E-10 & -- & 2.89E-11 & 6.40E-15 & 6.60E-16 & 3.12E-16 & 2.28E-16 & 8.16E-18 & 5.80E-18 & 4.13E-19\\
5s $^1$S & 3.16E-21 & 2.29E-14 & 2.85E-11 & 1.82E-09 & -- & 2.50E-12 & 2.31E-13 & 1.06E-13 & 7.61E-14 & 2.69E-15 & 2.00E-15 & 1.08E-16\\
5p $^3$P$^{\circ}$ & 1.74E-21 & 7.39E-15 & 5.03E-12 & 1.30E-10 & 8.06E-10 & -- & 1.18E-10 & 4.81E-11 & 3.28E-11 & 1.12E-12 & 9.10E-13 & 4.91E-14\\
4d $^1$D & 2.10E-21 & 8.25E-15 & 5.21E-12 & 1.23E-10 & 6.79E-10 & 1.07E-09 & -- & 2.88E-10 & 1.72E-10 & 4.24E-12 & 3.38E-12 & 2.00E-13\\
4d $^3$D & 5.62E-22 & 2.16E-15 & 1.33E-12 & 3.05E-11 & 1.64E-10 & 2.31E-10 & 1.52E-10 & -- & 1.04E-10 & 1.63E-12 & 1.28E-12 & 7.74E-14\\
5p $^1$P$^{\circ}$ & 2.59E-21 & 9.80E-15 & 5.97E-12 & 1.35E-10 & 7.15E-10 & 9.55E-10 & 5.50E-10 & 6.32E-10 & -- & 8.89E-12 & 6.99E-12 & 4.21E-13\\
6s $^3$S & 1.78E-21 & 1.08E-14 & 7.71E-12 & 1.82E-10 & 9.51E-10 & 1.23E-09 & 5.10E-10 & 3.71E-10 & 3.35E-10 & -- & 1.22E-12 & 6.08E-14\\
6s $^1$S & 6.16E-21 & 4.44E-14 & 3.61E-11 & 9.28E-10 & 5.06E-09 & 7.15E-09 & 2.92E-09 & 2.10E-09 & 1.89E-09 & 8.74E-12 & -- & 2.77E-14\\
Zn$^+$ + H$^-$ & 7.42E-19 & 1.67E-12 & 7.34E-10 & 1.26E-08 & 5.24E-08 & 7.37E-08 & 3.29E-08 & 2.41E-08 & 2.17E-08 & 8.32E-11 & 5.30E-12 & --\\
 \multicolumn{13}{l}{T = 2000 K} \\
  \multicolumn{13}{l}{...} \\
			\hline
\end{tabular}
\end{center}			
This table is available in its entirety in a machine-readable form in the online journal. A portion is shown here for guidance regarding its form and content.	
\end{table*}

Non-adiabatic nuclear dynamics is investigated using the analytical multichannel formulas \citep[see, e.g.][]{Belyaev-etc-aa-2014, Yakovleva-etc-aa-2016} based on the Landau-Zener model for individual non-adiabatic transitions. Knowing the state-to-state transition probabilities, the inelastic processes cross sections and rate coefficients are calculated by the standard formulas.

The detailed formulae for calculations of the cross sections and the rate coefficients are written for example in \citet[][]{Belyaev-etc-aa-2014, Yakovleva-etc-aa-2016}. The rate coefficients have the units of cm$^3$~s$^{-1}$ and the physical meaning of the product of velocities and cross sections averaged over the Maxwellian distribution. 
Being a result of multiplication of the rate coefficient defined above by the concentration of collided partners, the collision rates are entered in the SE equations and determine populations of atomic states.

\begin{figure}
	\includegraphics[width=\columnwidth]{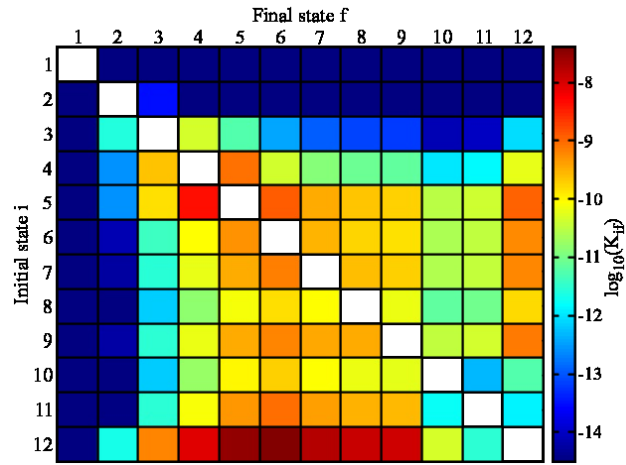}
	\caption{The graphical representation of the rate coefficients for excitation, de-excitation, neutralization and ion-pair formation processes in ${\rm Zn} + {\rm H}$ and ${\rm Zn}^+ + {\rm H}^-$ collisions at temperature $T = 6000$ K. 
		Initial and final state labels are presented in Table~\ref{tab:states}.
	}
	\label{fig:heatmap}
\end{figure}

The derived rate coefficients are presented in  Table~\ref{tab:rates}, and the graphical representation of the calculated rate coefficients at the temperature T = 6000 K is given in Fig.~\ref{fig:heatmap}. Rate coefficients are plotted with color according to their values, the study does not treat elastic processes and they are plotted by white. Fig.~\ref{fig:heatmap} shows that the highest values of the rate coefficients correspond to several partial mutual neutralization processes to the final scattering channels ${\rm Zn}(3d^{10}4s5s~^1S)$, ${\rm Zn}(3d^{10}4s5p~^3P^{\circ})$, ${\rm Zn}(3d^{10}4s4d~^{1,3}D)$, ${\rm Zn}(3d^{10}4s5p~^1P^{\circ}) + {\rm H}(1s~^2S_{\rfrac{1}{2}})$ (transitions 18 $\to$ 5 -- 9).

It is worth emphasizing that nowadays it is well recognized that the so-called classical Drawin formula \citep{Drawin1968,Drawin1969,Steenbock1984} widely used in the past for
estimating H-collision rate coefficients does not have a correct physical background and therefore does not yield reliable rate
coefficients. Moreover, it has been shown in \citet{2011A&A...530A..94B} that for optically allowed transitions the Drawin formula can overestimate rate coefficients up to a factor of several orders of magnitude, while it can underestimate optically forbidden transitions by several orders of magnitude. This means that no "correction factor" can be applied to rate coefficients obtained by the Drawin formula in order to get correct absolute values of rate coefficients, as well as to get correct ratios of rate coefficients, which are very important for non-LTE stellar atmosphere modeling. The proper way is to calculate a corresponding rate coefficient for each inelastic transition, as it is done in the present paper.

\subsection{Programs and model atmospheres}

The coupled radiative transfer and statistical equilibrium equations were solved with a revised version of the  \textsc {detail} code  \citep{detail}. The opacity package was updated for applications of the \textsc {detail} code to FGK and BA stars. The updates were presented by \citet{mash_fe} and \citet{2011JPhCS.328a2015P}, respectively.

For synthetic spectra calculations, we use the \textsc {synthV\_NLTE} code \citep{Tsymbal2018} mounted to the \textsc {idl binmag} code by \citet{2018ascl.soft05015K}\footnote{https://www.astro.uu.se/$\sim$oleg/binmag.html}. This technique allows to obtain the best fit to the observed line profiles with the NLTE effects taken into account via  pre-calculated departure coefficients for a given model atmosphere.

For late type stars, we used classical plane-parallel model atmospheres from the \textsc {marcs} model grid \citep{marcs}, which were interpolated for given \teff, \lgg, and [Fe/H].
For BA stars, the model atmospheres were calculated under the LTE assumption with the code \textsc {LLmodels} \citep{LLmodels}.

The lines of Zn\ione\ and Zn\ii\ selected for model atom testing and abundance determination are listed in Table~\ref{tab_lines}. For the majority of the selected lines, their oscillator strengths originate from laboratory measurements. An exception is Zn\ione\ 6362~\AA\ line, with calculated log~gf = 0.14 \citep{1980A&A....84..361B}. We note that, for the Zn\ione\ 4680, 4722 and 4810 \AA\ lines, calculated log~gf values are consistent within 0.02 dex with the measurements of \citet{2012ApJ...750...76R}. For Zn\ii\ 2062 \AA, measurements of \citet{2006EPJD...37..181M} and \citet{2012ApJ...750...76R} provide consistent within 0.01~dex log~gf.

For lines of Zn\ione,  Van der Waals damping constants ($\gamma_{\rm VdW}$) were calculated using the code of \citet{1998PASA...15..336B} and data from \citet{1995MNRAS.276..859A} and \citet{1997MNRAS.290..102B}, while, for Zn\ii, we adopted a standard formula from \citet{2005oasp.book.....G} used in the \textsc {synthV\_NLTE} code by default. 
Radiative damping constant ($\gamma_{\rm Rad}$) for the resonance lines of Zn\ione\ and Zn\ii\ were calculated from their oscillator strengths, while classical damping constants were adopted for the subordinate lines.
Stark damping constants ($\gamma_{\rm Stark}$) for Zn\ione\ lines were calculated with an approximate formula of \citet{1971Obs....91..139C}. In cool stars, Stark effect is negligible. In hot stars, Stark effect is larger, but Zn\ione\ lines in the visible range are weak. For the lines of Zn\ii, $\gamma_{\rm Stark}$ were taken from \citet{BLb}.
 The adopted damping constants are presented in Table~\ref{tab_lines}.

\begin{table*}
	\caption{The list of lines}
	\label{tab_lines}
	\setlength{\tabcolsep}{0.9mm}
	\begin{tabular}{lrclrlccc}
		\hline
		Ion  & $\lambda$, \AA &  transition & \eexc, eV  & log gf & ref. & log$\gamma_{\rm Rad}$ & log$\gamma_{\rm Stark}$ & log$\gamma_{\rm VdW}$  \\
		\hline
 Zn I  &   2138.573 &  4s$^1$S  --  4p$^1$P$^{\circ}$ & 0    & 0.16   & A75 & 8.85 & --5.47 & --7.64  \\ 
 Zn I  &   3075.895 &  4s$^1$S  --  4p$^3$P$^{\circ}$ & 0    & --3.85 & R12 & 4.58 & --5.82 & --7.86  \\
 Zn I  &   3302.583 &  4p$^3$P$^{\circ}$  --  4d$^3$D & 4.03 & --0.02 & R12 & 8.31 & --4.78 & --7.25  \\
 Zn I  &   4680.136 &  4p$^3$P$^{\circ}$  --  5s$^3$S & 4.01 & --0.85 & R12 & 8.01 & --5.24 & --7.30  \\
 Zn I  &   4722.153 &  4p$^3$P$^{\circ}$  --  5s$^3$S & 4.03 & --0.37 & R12 & 8.00 & --5.24 & --7.30  \\
 Zn I  &   4810.528 &  4p$^3$P$^{\circ}$  --  5s$^3$S & 4.08 & --0.15 & R12 & 7.98 & --5.24 & --7.30  \\
 Zn I  &   6362.338 &  4p$^1$P$^{\circ}$  --  4d$^1$D & 5.80 &  0.14 & BG80 & 7.74 & --4.80 & --7.30  \\
 Zn II &   2025.484 &  4s$^2$S  --  4p$^2$P$^{\circ}$ & 0  & --0.03 & M06 & 8.61 & --6.67 & --7.90  \\
 Zn II &   2062.001 &  4s$^2$S  --  4p$^2$P$^{\circ}$ & 0  & --0.30 & M06 & 8.59 & --6.72 & --7.90  \\
		\hline
\end{tabular}\\
A75 = \citet{1975JQSRT..15...25A}, R12 = \citet{2012ApJ...750...76R}, BG80 = \citet{1980A&A....84..361B}, M06 = \citet{2006EPJD...37..181M} \\
\end{table*}

\section{Statistical equilibrium of Zn\ione--\ii}\label{se}

\begin{figure}
	\includegraphics[width=80mm]{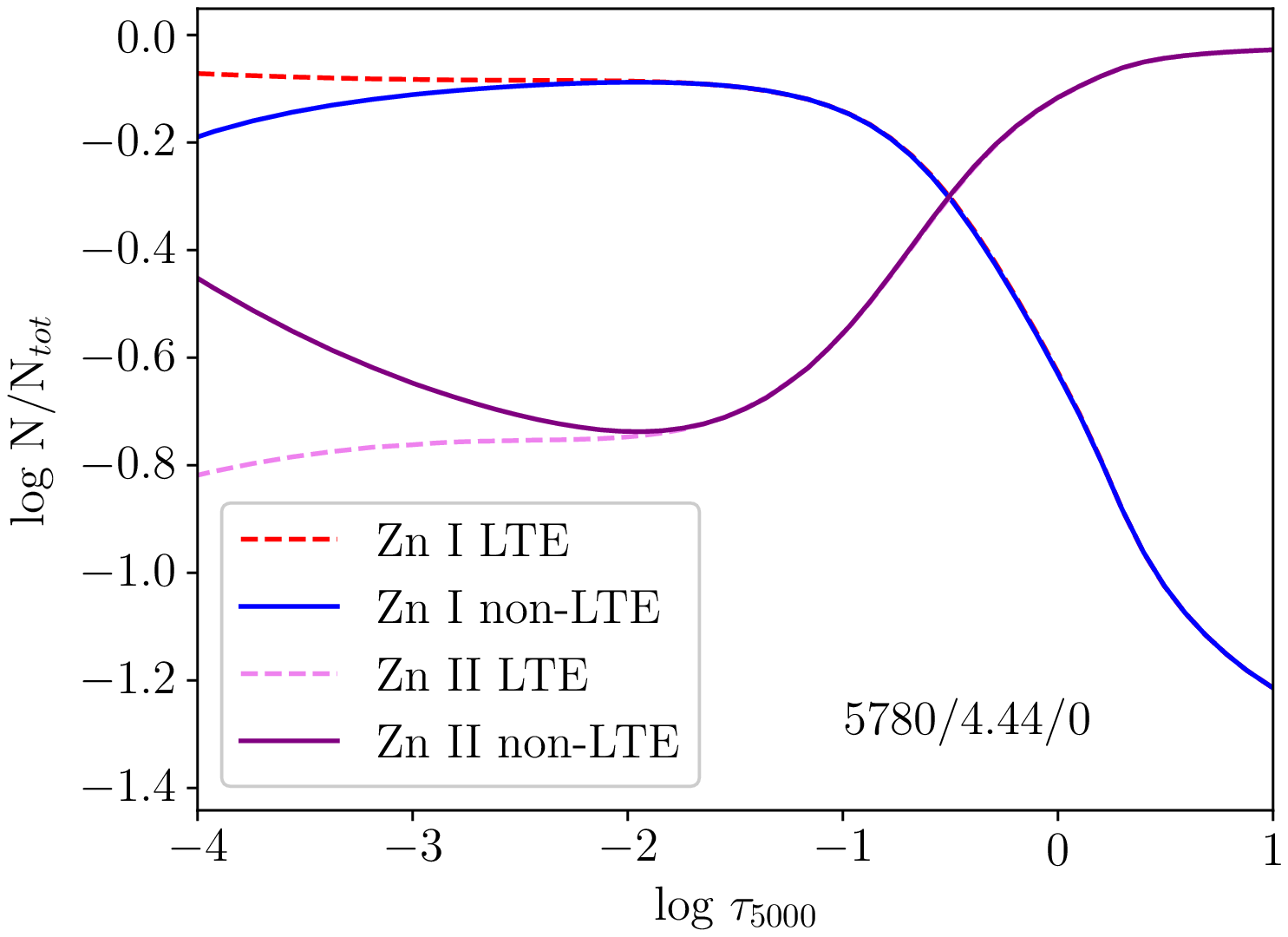}
	\includegraphics[width=80mm]{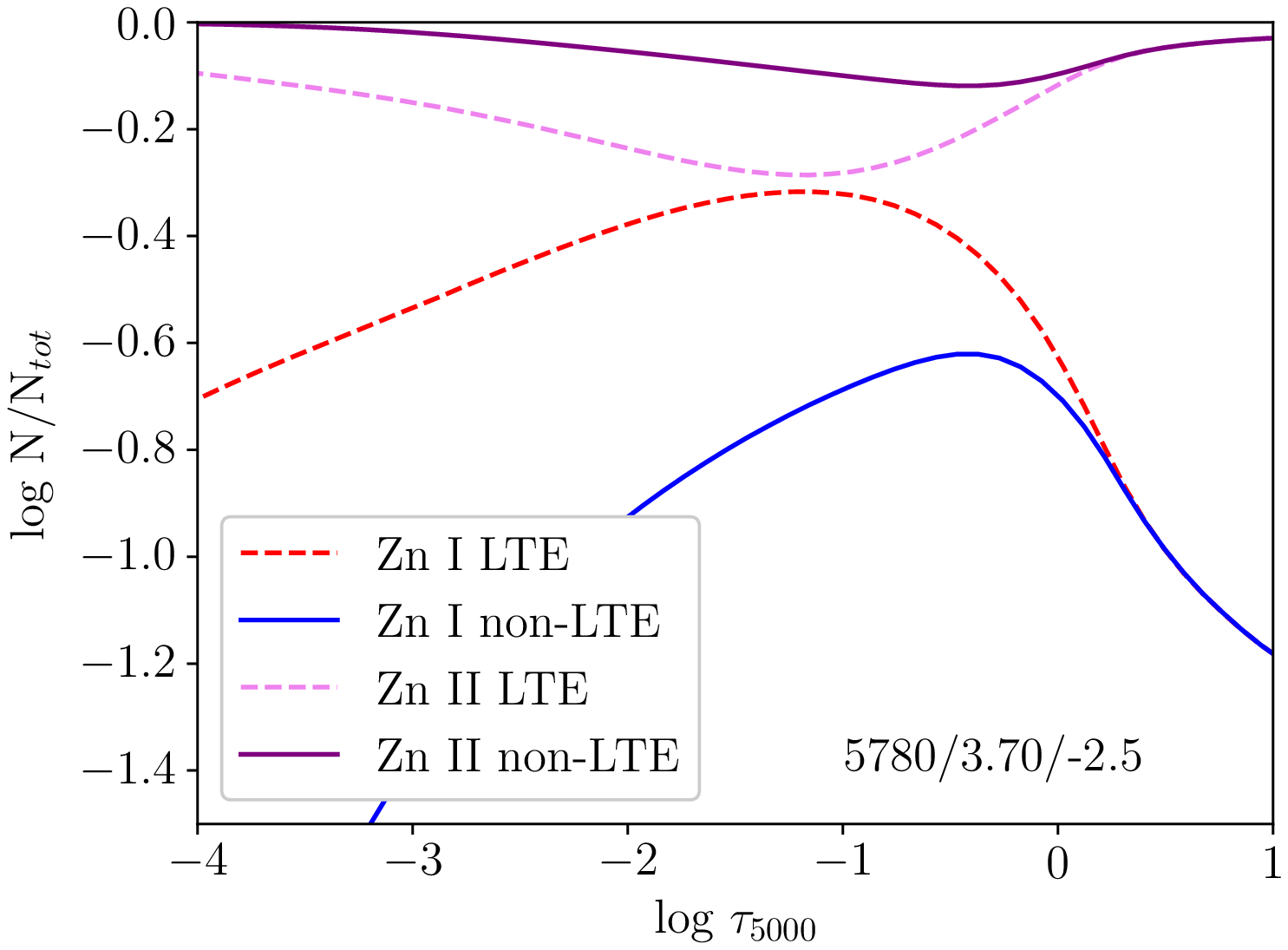}
	\includegraphics[width=80mm]{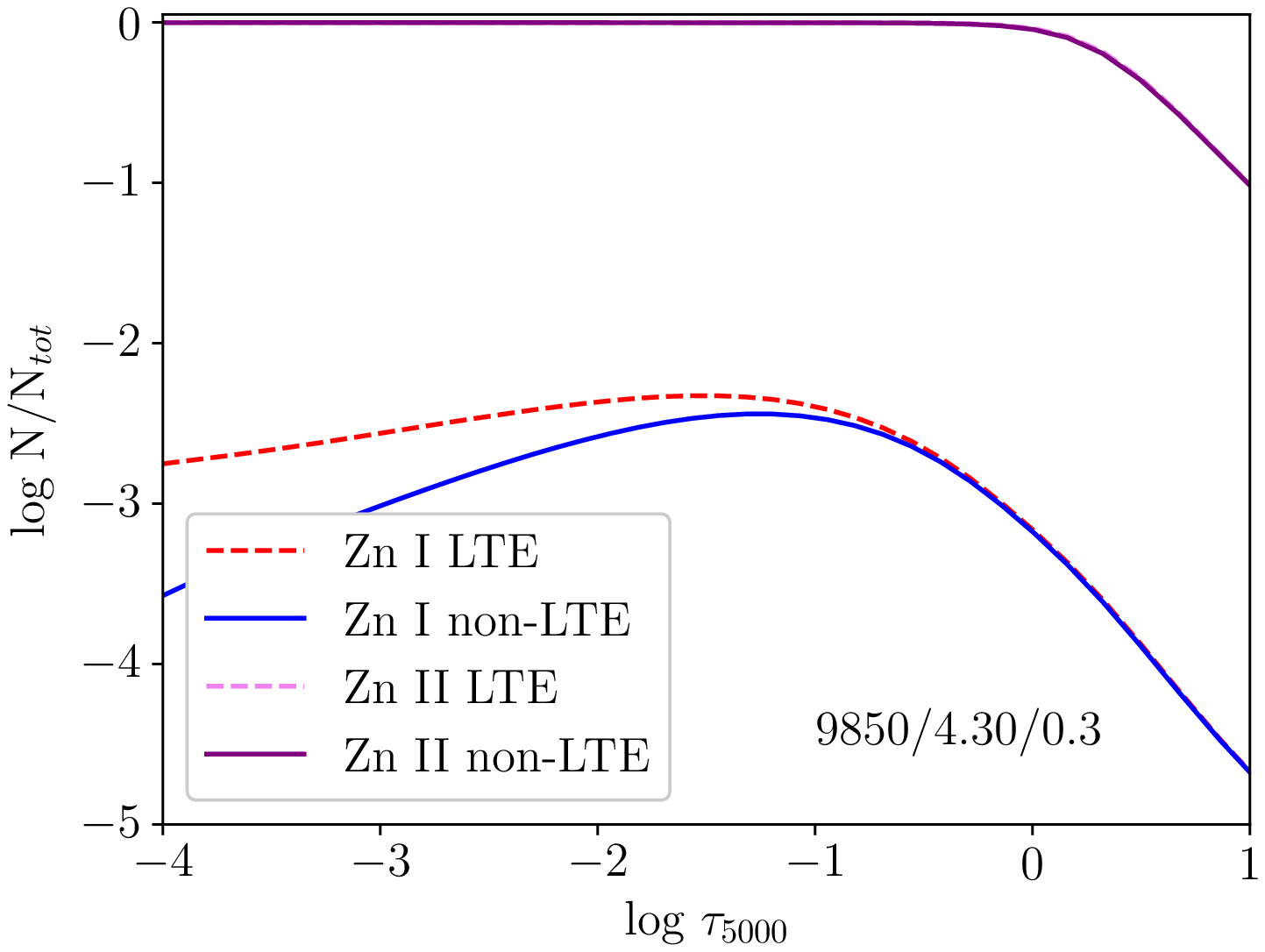}	
	\caption{	
		Non-LTE and LTE number densities of Zn\ione\ and Zn\ii\ in model atmosphere of Sun (top panel), VMP star HD~140283 (middle panel), and Sirius (bottom panel). For each model, \teff/log~g/[Fe/H] are indicated.}
	\label{ions} 
\end{figure}

The ionisation threshold of neutral zinc is 9.39~eV and, in atmospheres of GK-spectral type stars, number densities of neutral and ionised zinc take values of nearly the same order of magnitude. In atmospheres of BAF-type stars, Zn\ii\ dominates.  Fig.~\ref{ions} shows non-LTE and LTE number densities of Zn\ione\ and Zn\ii\ in model atmospheres with different stellar parameters, which represent Sun, very metal-poor (VMP) star HD~140283, and Sirius. 

\begin{figure}
	\includegraphics[width=80mm]{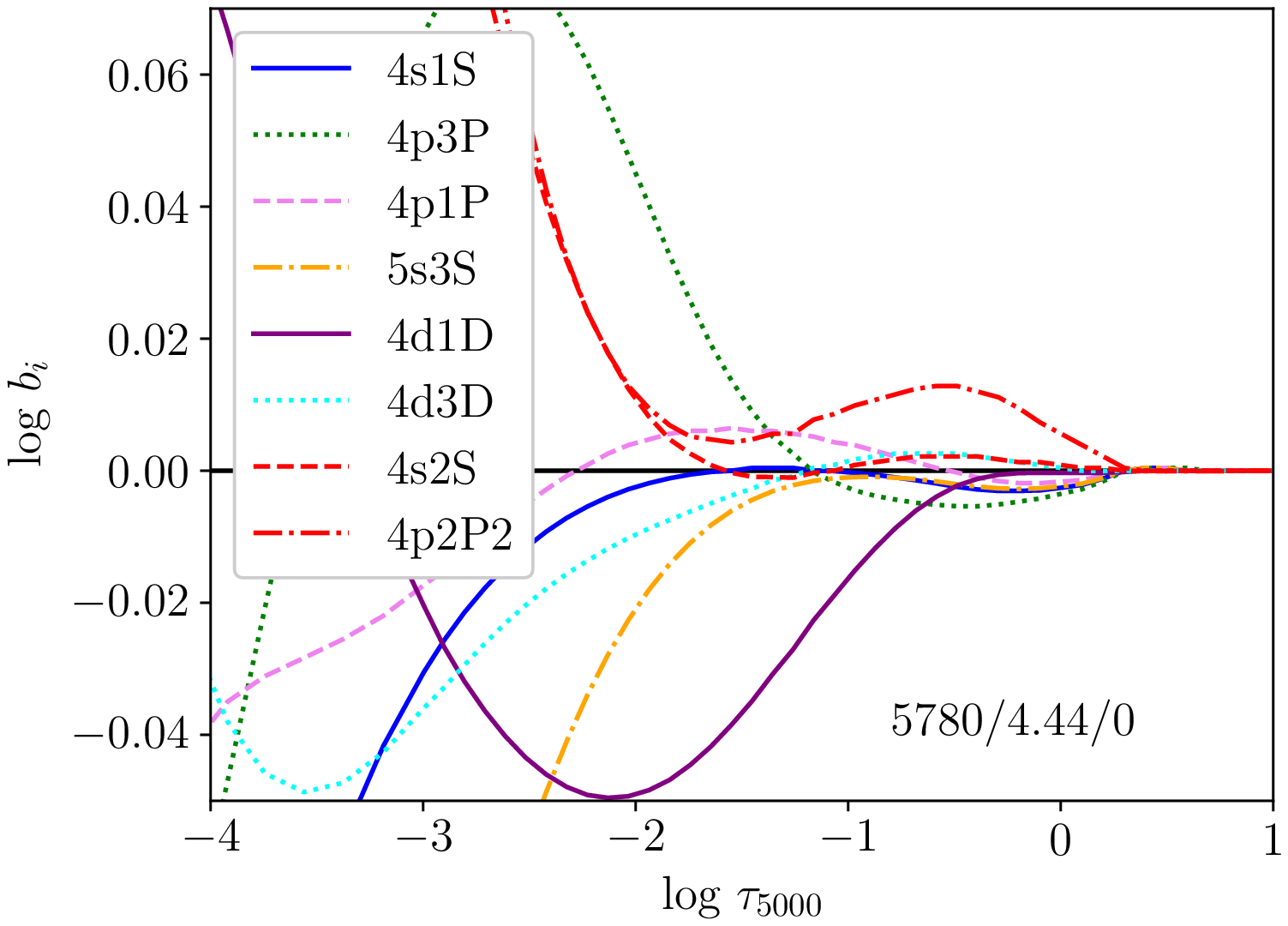}
	\includegraphics[width=80mm]{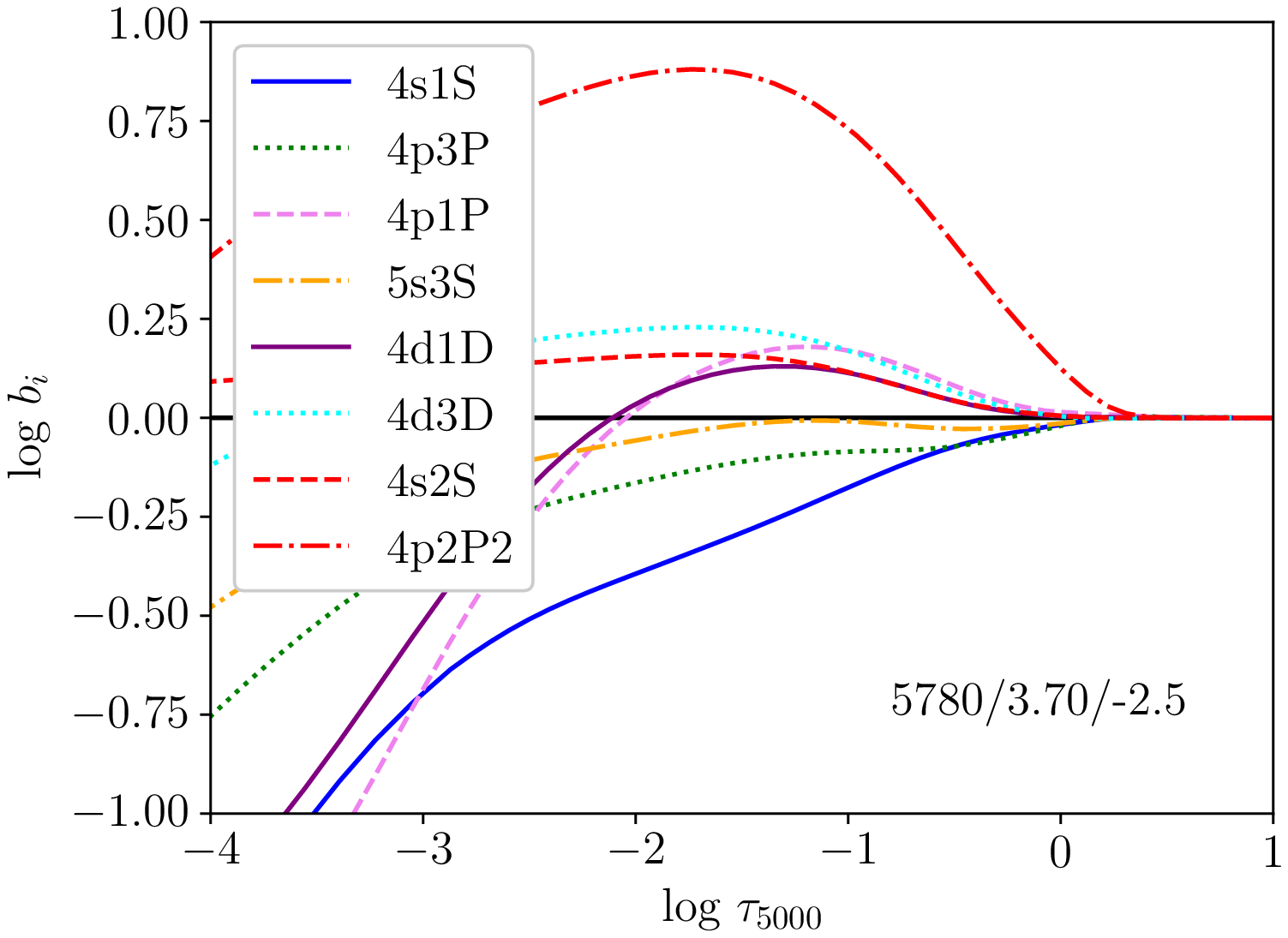}
	\includegraphics[width=80mm]{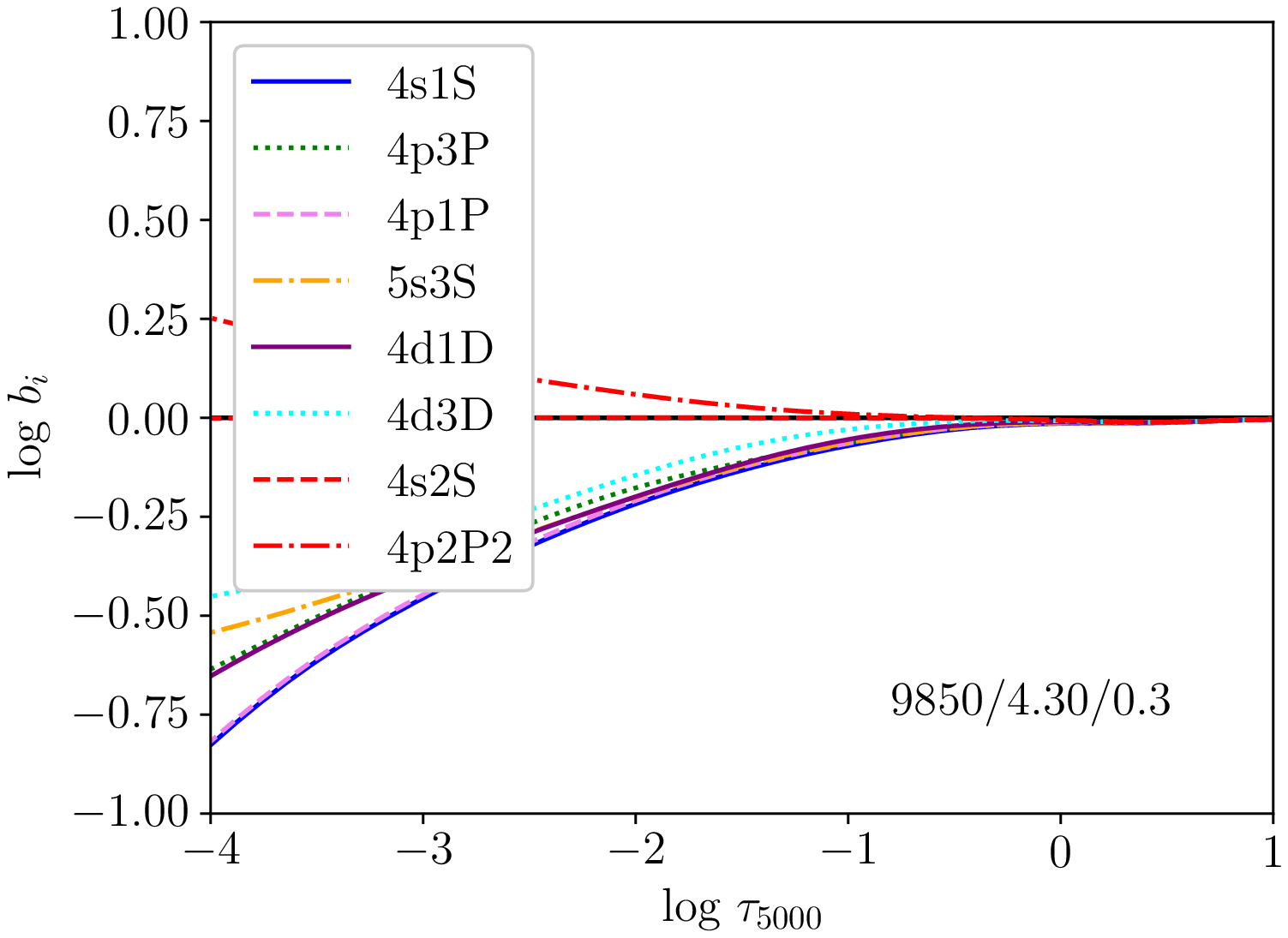}	
	\caption{	
		Departure coefficients for selected atomic levels of Zn\ione\ and Zn\ii\ in model atmosphere of Sun (top panel), VMP star HD~140283 (middle panel), and Sirius (bottom panel). For each model, \teff/log~g/[Fe/H] are indicated.}
	\label{bfs} 
\end{figure}

The deviations from LTE are usually characterised with departure coefficients b$_i$ = n$_{\rm i, NLTE}$/n$_{\rm i, LTE}$, where n$_{\rm i, NLTE}$ and n$_{\rm i, LTE}$ are non-LTE and LTE populations of $i$ atomic level.
Spectral line profiles are affected  by  non-LTE effects  via the changes in the line opacity, which depends mainly on b$_i$, and the line source function, which mainly depends on the ratio of the departure coefficient of the upper and lower level (b$_j$/b$_i$).
The departure coefficients for selected atomic levels of Zn\ione\ and Zn\ii\  in model atmospheres of the Sun, VMP star HD~140283, and Sirius are shown in Fig.~\ref{bfs}. The investigated lines mostly form in layers with optical depth at $\lambda$ = 5000~\AA, log~$\tau_{\rm 5000} > -2.7$.  

In deep atmospheric layers with log~$\tau_{\rm 5000} > 0.5$, collisions dominate, the departures from LTE are missing and the atomic levels keep their LTE populations. Non-LTE effects become more prominent towards higher atmospheric layers. When studying model atmospheres with different stellar parameters, generally, the non-LTE effects increase with increasing effective temperature and decreasing surface gravity and metallicity. 

In solar atmosphere, Zn\ione\ slightly prevails upon Zn\ii\ in atmospheric layers with log~$\tau_{\rm 5000} < -0.5$, while, in deeper layers,  the opposite situation arises (Fig.~\ref{ions}).  This results in a minor depopulation of the ground state of Zn\ione\ at log~$\tau_{\rm 5000} \simeq -0.5$. Although the model atmosphere of the VMP star has close to solar \teff, the ground state of Zn\ione\  is strongly depopulated in HD~140283 (Fig.~\ref{bfs}). This happens due to smaller electron number density, stronger ionisation degree, and, thus, stronger overionisation of Zn\ione. 

In FGK stars, the deviations from LTE for Zn\ione\ are mainly driven by bound-free transitions in metal-poor stars, while, in stars with close to solar [Fe/H], bound-bound transitions are also important. 
For Zn\ione\ 4680, 4722, 4810 \AA\ lines non-LTE may lead to either stronger or weaker lines depending on stellar parameters.
In solar atmosphere, the line source function is smaller than the Planck function, which results in strengthened lines and negative abundance corrections.
The low excitation levels 4p$^3$P$^{\circ}$ and 4p$^1$P$^{\circ}$ are less depopulated compared to the ground state due to the radiative pumping transitions from the ground state. 
In stars with higher \teff\ and/or lower [Fe/H],  these lines are weak and, in contrast,  become even weaker in non-LTE.
In cool metal-poor stars, strong underpopulation of the ground state always results in weakened Zn\ione\  lines and positive non-LTE abundance corrections.   

The levels of Zn\ii\  are overpopulated due to depopulation of  Zn\ione\ levels. While in solar atmosphere this effect is small, it increases with decreasing metallicity and becomes prominent in metal-poor stars (see middle panels of Fig.~\ref{bfs} and also Fig.~\ref{ions}). An overpopulation of the ground state results in the line opacity increase and tends to strengthen resonance  lines of Zn\ii.  However, the  4p$^2$P$^{\circ}$ level is more overpopulated compared to the ground state of Zn\ii. This results  in an increase in the line source function and tends to weaken resonance lines of Zn\ii. Which effect prevails depends on stellar parameters and the line strength. For the resonance lines of Zn\ii, non-LTE may lead to either positive or small in absolute value negative abundance corrections.

\section{Testing non-LTE method with the reference stars}\label{testing}

We test our non-LTE abundance determination method with the Sun and three metal-poor stars with well-determined atmospheric parameters (Table~\ref{tab_abun}). 

\subsection{Solar zinc abundance}

Solar abundance was determined using the spectrum of the Sun as a star \citep{kurucz84}.
The classical 1D  model atmosphere with  effective temperature \teff\ = 5780~K, surface gravity log~g = 4.44, and microturbulent velocity \vt\ = 0.9 \kms\ was taken from  the \textsc {marcs} webpage\footnote{https://marcs.astro.uu.se/}.

Our solar zinc abundance relies on three lines of Zn\ione, namely, Zn\ione\ 4722, 4810, and 6362 \AA. We found an average log~A$_\odot$\footnote{Here log~A(X) = log($N_X/N_{tot}$), where N$_{tot}$ is a total number density; X~I--X~II means difference in abundance derived from lines of X~I and X~II, log~A(X~I)--log~A(X~II). The abundances can be transformed to the scale relatively to the number of H atoms: $\eps$ = log~A + 12.04}. = $-7.40 \pm 0.04$ and $-7.45 \pm 0.02$ in LTE and non-LTE, respectively. The abundance error is calculated as the dispersion in the single line measurements around the mean
$\sigma = \sqrt{ \Sigma (log~A - log~A_i )^2 /(N - 1)}$, where N is a total number of lines. For the Sun, non-LTE leads to lower zinc abundance and better agreement between abundances from different lines.
Our non-LTE abundance is close to the meteoritic one log~A$_\odot$ = $-7.43$ \citep{2021SSRv..217...44L}.

We compare line-by-line solar LTE abundances and non-LTE abundance corrections with results of \citet{Takeda2005zn}. When using the same log~gf values as adopted by T05, for Zn\ione\ 4722, 4810, and 6362 \AA, we found log~A$_{\rm LTE}$ = $-7.40$, $-7.40$, and $-7.45$, respectively, while T05 found $-7.43$, $-7.50$, and $-7.51$ for the same lines. For Zn\ione\ 4722 and 4810 \AA\ lines, a difference in the adopted in this study \vt\ = 0.9 \kms\ and  1.0 \kms\ in T05 translates in abundance shift of 0.04~dex  and partially explains the discrepancy between the two studies. The remaining minor discrepancy can be caused by using different solar observed spectra, continuum placement and line fitting procedure. As for the non-LTE abundance correction, the agreement between the two studies is good, given that the departures from LTE are small in the solar atmosphere and different atomic data were used in the non-LTE calculations. For Zn\ione\ 4722, 4810, and 6362 \AA, we found $\Delta_{\rm NLTE}$ = $-0.06$, $-0.08$, and $-0.02$, and T05 gives $-0.05$, $-0.05$, and $0.0$, respectively. 

\subsection{Zn\ione\ and Zn\ii\ lines in  VMP stars}

For our analysis, we selected three VMP stars with well determined stellar parameters. 
Their effective temperatures are based on multiple photometric determinations and surface gravities -- on accurate distances. 
These VMP stars are  turn-off star HD~84937 with \teff/log~g/[Fe/H]/\vt = 6350/4.09/--2.18/1.7, subgiant HD~140283 (5780/3.70/--2.43/1.3), and giant HD~122563 (4600/1.40/--2.55/1.6).
They served as the reference stars in a number of non-LTE studies \citep[for example,][]{mash_fe,2019A&A...631A..43M,2011MNRAS.413.2184B,2017ApJ...847...15B,sitnova_ti}.  

We adopted high resolution and high signal to noise ratio spectra in the visible range from the ESO UVESPOP survey \citep{Bagnulo2003} and UV  spectra derived at Space Telescope Imaging Spectrograph  (STIS HST) from the StarCAT catalogue of Thomas Ayres\footnote{https://casa.colorado.edu/$\sim$ayres/StarCAT}. 

For test calculations, we determined abundances from different lines of Zn\ione\ (2138, 3075, 3302, 4810 \AA) and Zn\ii\ (2025, 2062 \AA) in HD~84937 and HD~140283 using different line formation scenarios:\\
-- LTE;\\
-- Non-LTE with electron collision rates calculated with approximate formulae as described in Sect.~\ref{atom} and neglecting inelastic collisions with hydrogen atoms;\\
-- Non-LTE with electron collision rates calculated with approximate formulae and quantum-mechanical rate coefficients for hydrogenic collisions;\\
-- Non-LTE with accurate quantum-mechanical data for electronic and hydrogenic collisions.

\begin{figure}
	\includegraphics[width=80mm]{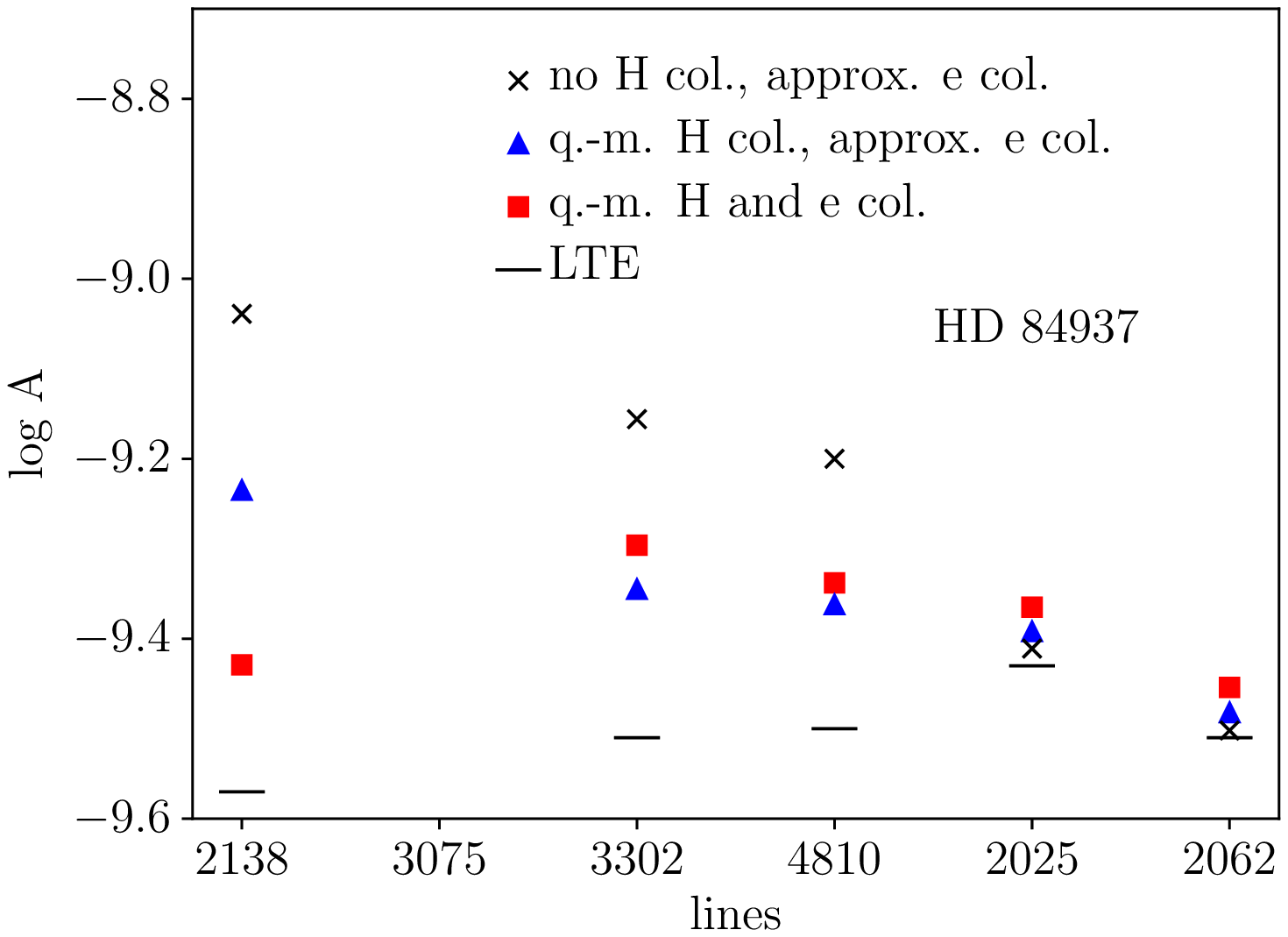}
	\includegraphics[width=80mm]{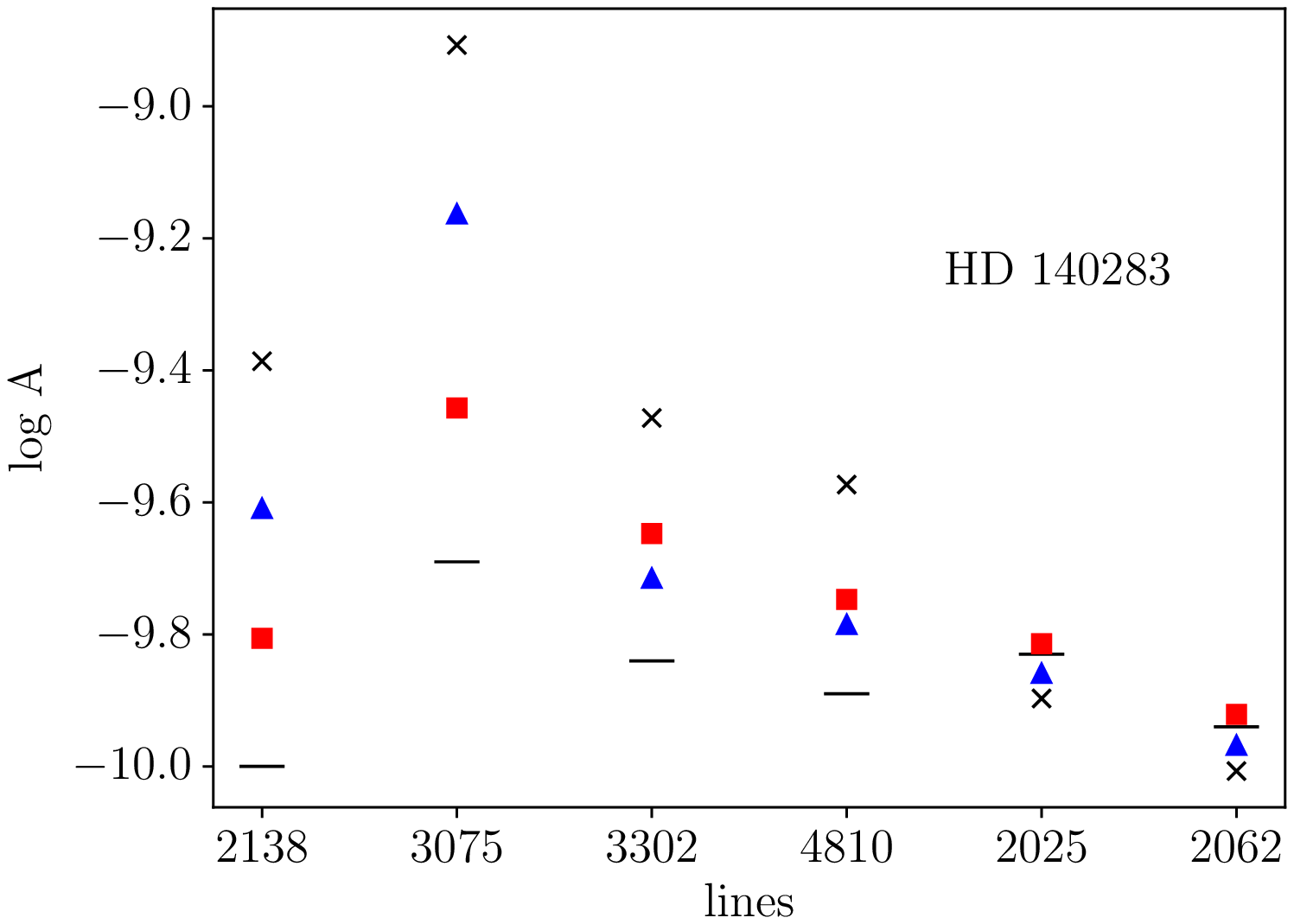}
	\caption{	
		Zinc abundances from different lines of Zn\ione\ (2138, 3075, 3302, 4810 \AA) and Zn\ii\ (2025, 2062 \AA) in HD~84973 (top panel) and HD~140283 (bottom panel) at different line formation scenarios: LTE (dashes), non-LTE with neglecting H collisions and approximate electronic collisions (crosses), non-LTE with quantum-mechanical H collisions and approximate electronic collisions (triangles), non-LTE with quantum-mechanical data for H and e collisions (squares). See text for details.}
\label{scenarios} 
\end{figure}

Fig.~\ref{scenarios} shows the abundances from individual lines in HD~84937 and HD~140283 derived in different line formation scenarios.
 In HD140283, the LTE abundance difference between any lines  does not exceed 0.3~dex. 
LTE abundances from the subordinate lines of Zn\ione\ agree well, while the resonance line Zn\ione\ 2138 \AA\ and the forbidden Zn\ione\ 3075 \AA\ line provides 0.15~dex lower and higher LTE abundance, respectively, compared to the other Zn\ione\ lines. 
The observed spectrum of HD~84937 does not cover the forbidden Zn\ione\ 3075 \AA\ line, and LTE abundances between different zinc lines agree within 0.07~dex. 

In non-LTE, when neglecting inelastic collisions with H atoms and using approximate electron collision rates, the departures from LTE are huge, and the non-LTE abundance correction for the resonance Zn\ione\ 2138 \AA\ line $\Delta_{\rm NLTE}$ =  0.53 and 0.61~dex in HD~84937 and HD~140283, respectively. For the subordinate Zn\ione\ 3302 and 4810 \AA\ lines, the corrections are smaller: 0.35 and 0.30 dex in HD~84937  and 0.37 and 0.32 dex in HD140283. 

In the investigated stars, Zn\ii\ is the majority species, and non-LTE abundance corrections for the Zn\ii\ resonance lines  do not exceed 0.08~dex in absolute value in any non-LTE line formation scenario. The Zn\ii\ 2025 and 2062 \AA\ resonance lines belong to the same multiplet, however, in HD~140283 and HD~84937, Zn\ii\ 2025 \AA\ provides 0.1~dex higher abundance compared to the Zn\ii\ 2062 \AA\ line, either in LTE or non-LTE. 
The abundance from Zn\ii\ 2025 \AA\ can be uncertain due to  blending with Fe\ione, Ni\ione, Cu\ii, and Cr\ii\ lines. 
When fitting the line profile, we varied abundances of the above elements, however, the continuum placement can be unreliable due to overlapping wings of the lines. We prefer to keep the abundance from this line as a sanity check due to the small number of zinc lines. 

Including quantum-mechanical rate coefficients for inelastic collisions with H atoms decreases the departures from LTE, which results in a decrease of 0.2~dex in $\Delta_{\rm NLTE}$ for all  lines Zn\ione\ of in the two stars. 

When using accurate electronic collision rates based on  R-matrix calculations of  \citet{2005PhRvA..71b2716Z},  the non-LTE effects  for the Zn\ione\ resonance line decreases even more, and   $\Delta_{\rm NLTE} (2138)$ =  0.19 and 0.14~dex in HD~140283 and HD~84937, respectively. For the subordinate lines Zn\ione\ 3302 and 4810 \AA, the changes are minor.

The latter non-LTE line formation scenario is our final one, and, for this case, the derived non-LTE abundances from different lines are presented in Table~\ref{tab_abun}. In this scenario, non-LTE abundances from Zn\ione\ 2138, 3302, and 4810 \AA\, are found to be consistent within 0.13~dex and 0.16~dex in HD~84937 and HD~140283, respectively. Either in LTE or in non-LTE, the abundance from the forbidden Zn\ione\ 3075  \AA\ line in HD~140283 is higher compared to the other lines. 

For HD~140283, we estimated an impact of uncertainties in \teff, log~g, and \vt\ on abundances from different lines (Table~\ref{tab_abun}). We adopted $\Delta$ \teff\ = 50~K \citep{2018MNRAS.475L..81K},   $\Delta$ log~g = 0.05~dex, and $\Delta$ \vt\ = 0.2 \kms\ \citep{2019A&A...631A..43M}. The above changes lead to abundance shifts of less than 0.08~dex in absolute value and cannot cancel the discrepancy between the Zn\ione\ 3075 \AA\ line and other lines.

For HD~122563, we determined zinc abundance from the forbidden Zn\ione\ 3075 \AA\ line and Zn\ione\ 4722 and 4810 \AA\ lines in LTE and non-LTE, using the latter scenario. The UV spectrum of this star does not cover Zn\ii\ lines.
The subordinate lines give very similar non-LTE abundances log~A = $-9.90$ and $-9.91$, while lower log~A = $-10.08$ was found from the forbidden line. This behaviour is different compared to the HD~140283. We do not recommend to adopt the forbidden Zn\ione\ 3075 \AA\ line for zinc abundance determination neither in non-LTE, nor in LTE.

As for the Zn\ione\ -- Zn\ii\ ionisation balance in HD~84937 and HD~140283, for each star, the average abundances from the two ionization stages agree within 0.08~dex either in LTE or non-LTE. For both stars, non-LTE leads to 0.17~dex higher abundance for Zn\ione, while, for Zn\ii, the non-LTE abundance is 0.06~dex and 0.03~dex higer ain HD~84937 and HD~140283, respectively. 
Here, we excluded the Zn\ione\ 3075 \AA\ line, when calculating the average abundance from Zn\ione. We found this line is an unreliable abundance indicator and we do not recommend to apply it in abundance analysis either in LTE or non-LTE.

We calculated the non-LTE abundance correction for the Zn\ione\ 2138 \AA\ line detected  by \citet{2019ApJ...876...97E} in the UV spectrum of hyper metal-poor star HE1327-2326. LTE zinc abundance determination together with non-LTE iron abundances results in [Zn/Fe] = 0.80 \citep{2019ApJ...876...97E}. The extremely low metallicity [Fe/H] = $-5.2$ together with high enough \teff\ = 6180~K and low log~g = 3.7 \citep{2005Natur.434..871F} results in dramatic departures from LTE, and $\Delta_{\rm NLTE}$ = 1.03~dex for  Zn\ione\ 2138 \AA\ line.

\section{BAF type stars}\label{baf}

We determine zinc abundance in a sample of BAF-spectral type stars with well-determined stellar parameters. For these stars, the non-LTE abundance determination was performed for a number of chemical elements, see \citet[][hereafter MR20]{2020MNRAS.499.3706M} for details. For the summary of stellar parameter determination and description of high quality observed spectra adopted for abundance determination of the BAF stars we refer the reader to MR20. 
We also determined zinc abundance in Procyon, which is missing in MR20. 
For Procyon, we adopted stellar parameters  from \citet{ryab2015}, and we refer a reader to this paper for details on atmospheric parameter determination and the description of the observed spectra.

For the BAF sample stars, we determine Zn abundance from the 4680, 4722, 4810 \AA\ lines. From one to three lines were adopted depending on their strength in the observed spectra. 
The derived LTE and non-LTE abundances are presented in Table~\ref{tab_abun1}.

The two coolest stars, HD 32115 and Procyon, are deficient in Zn in both LTE and non-LTE calculations, like the close-to-solar metallicity G dwarfs studied in Sect.~\ref{znfe}.

In contrast, the hotter A and late B-type stars are enhanced in Zn already in LTE and even more in non-LTE. The two of them, Sirius and HD 72660, are known as metallic-line (Am) stars and reveal a steep growth of abundances of the heavy elements beyond Fe (MR20, their Fig. 4). Their high values of [Zn/H] $\simeq$ 1 lie well in their element abundance patterns.

The remaining three stars, HD~145788, HD~73666, and 21~Peg, reveal similar overabundances of Zn relative to Fe, with [Zn/Fe]$_{\rm non-LTE}$ = 0.22, 0.29, and 0.34, respectively. They have different metallicities because formed in the environments of different chemical composition, but very similar element abundance patterns, which resemble that for the superficially normal stars (see MR20 for a discussion and references). As found by MR20, in the superficially normal stars, the non-LTE abundances of the chemical elements from carbon to iron are consistent within 0.1 dex with either solar (for the Galactic field stars) or scaled solar (for the star HD~73666 in the Praesepe open cluster) abundances, however, the heavy elements Sr, Zr, Ba, and Nd reveal supersolar abundances. For example, in our hottest and Galactic field star 21~Peg, [Sr/H] = 0.59, [Zr/H] = 0.35,[Ba/H] = 0.98, and [Nd/H] = 0.48. The obtained high abundances of Zn support the abundance trends found by MR20.  
 





\begin{table}
	\caption{LTE and non-LTE line-by-line zinc abundances in the reference stars}
	\label{tab_abun}
	\setlength{\tabcolsep}{1.10mm}
\renewcommand{\arraystretch}{0.9} 
	\begin{tabular}{lrrrrrrrr}
		\hline
	& \multicolumn{6}{c}{Zn\ione} & \multicolumn{2}{c}{Zn\ii} \\
Lines:	 & 2138 & 3075 & 3302 & 4722 & 4810 & 6362 & 2025$^b$ & 2062 \\
		\hline
		\multicolumn{9}{c}{Sun, 5780/4.44/0.0/0.9} \\
 LTE       & -- &  -- & -- & --7.37  & --7.39  & --7.44 & -- & -- \\ 
 NLTE      & -- &  -- & -- & --7.43  & --7.47  & --7.46 & -- & -- \\ 
\multicolumn{9}{c}{HD~122563, 4600/1.43/--2.55/1.6} \\
LTE       &  --  &  --10.14   & -- &  --10.04  &  --10.04 & -- & -- & -- \\ 
NLTE      &  --  &  --10.08   & -- &  --9.90  & --9.91  & -- & -- & -- \\ 
		\multicolumn{9}{c}{HD~84937, 6350/4.09/--2.12/1.7} \\
 LTE       & --9.57   &  -- & --9.51 &  --  & --9.50  & -- & --9.43 & --9.51 \\  
 NLTE      & --9.43 & -- & --9.30 & -- & --9.34 & -- & --9.37 & --9.45 \\
		\multicolumn{9}{c}{HD~140283, 5780/3.70/--2.46/1.6} \\
 LTE       & --10.00  &  --9.69 & --9.84 &  --  & --9.89  & -- & --9.83 & --9.94 \\ 
 NLTE      & --9.81 & --9.46 & --9.65 & --& --9.75 & -- & --9.81 & --9.92 \\
 \hline
\multicolumn{9}{c}{Abundance shifts caused by changes in parameters of HD~140283} \\
 5730 & --0.05  & --0.05    & --0.02  &  --   & --0.02   & -- &   0.05 &   0.02 \\
 3.65 &   0.01  & --0.01    & --0.01  &  --   & --0.01   & -- & --0.03 & --0.04  \\ 
 1.4 &    0.01  &   0.01    &   0.00  &  --   &   0.00   & -- &   0.07 &   0.08 \\
		\hline
	\end{tabular}
$^b$ - Zn\ii\ 2025 \AA\ is blended and should be considered with a caution. 
\end{table} 

\begin{table}
	\caption{Atmospheric parameters and the average [Zn/H] in BAF sample stars}
	\label{tab_abun1}
	\setlength{\tabcolsep}{0.5mm}
\renewcommand{\arraystretch}{0.9} 
	\begin{tabular}{lrrrrrcc}
		\hline
Star	 & \teff & log~g & [Fe/H] & \vt & N$_l$ & [Zn/H]$_{\rm LTE}$ & [Zn/H]$_{\rm NLTE}$ \\
		\hline
		Procyon &  6590 & 3.95 & --0.02 & 1.8 &   3 & --0.15 $\pm$   0.08 & --0.13 $\pm$   0.10 \\ 
		HD~32115 &  7250 & 4.20 & 0.09 & 2.3 &    2 & --0.38 $\pm$   0.01 & --0.25 $\pm$   0.01 \\ 
		HD~73666 &  9380 & 3.78 & 0.24 & 1.8 &    2 &  0.34 $\pm$   0.01  & 0.53 $\pm$   0.01 \\ 
		HD~72660 &  9700 & 4.09 & 0.67 & 1.8 &    3 &  0.87 $\pm$   0.04  & 1.02 $\pm$   0.04 \\ 
		HD~145788 &  9750 & 3.70 & 0.52 & 1.3 &    1 &  0.55   & 0.74  \\ 
		Sirius &  9850 & 4.30 & 0.30 & 1.8 &    3 &  0.84 $\pm$   0.02  & 0.99 $\pm$   0.02 \\ 
		21 Peg &  10400 & 3.55 & 0.05 & 0.5 &    1 &  0.21 &  0.39 \\ 
		\hline
	\end{tabular}
\end{table}

\section{Zinc abundance in the sample of FGK dwarfs in a wide metallicity range}\label{znfe}

The FGK sample stars are listed in Table~\ref{tab_znh} together with their stellar atmosphere parameters taken from our previous studies \citep{lick,2019ARep...63..726M,2003A&A...397..275M}. 
The stars cover the $-2.5 \leq$ [Fe/H] $\leq 0.2$ metallicity range. \citet{lick} and \citet{2019ARep...63..726M} determined atmospheric parameters with a common method, which is based on photometric \teff\ and distance-based log~g (see the above papers for the details). \citet{2003A&A...397..275M} also determined distance-based log~g, while \teff\ was derived from fitting the Balmer line wings.

High-resolution ($\lambda$/$\Delta$$\lambda >$ 45~000)
spectra with a signal-to-noise ratio S/N $>$ 60 were
obtained at the 3-m telescope of the Lick Observatory
with the Hamilton spectrograph or taken from
the UVES\footnote{http://archive.eso.org/eso/eso\_archive\_main.html} and ESPaDOnS\footnote{http://www.cadc-ccda.hia-iha.nrc-cnrc.gc.ca/en/search/} archives. We also
used the spectra obtained at the 2.2-m telescope
of the Calar Alto Observatory with the FOCES
spectrograph and provided by K. Fuhrmann. 

\begin{figure}
	\includegraphics[width=80mm]{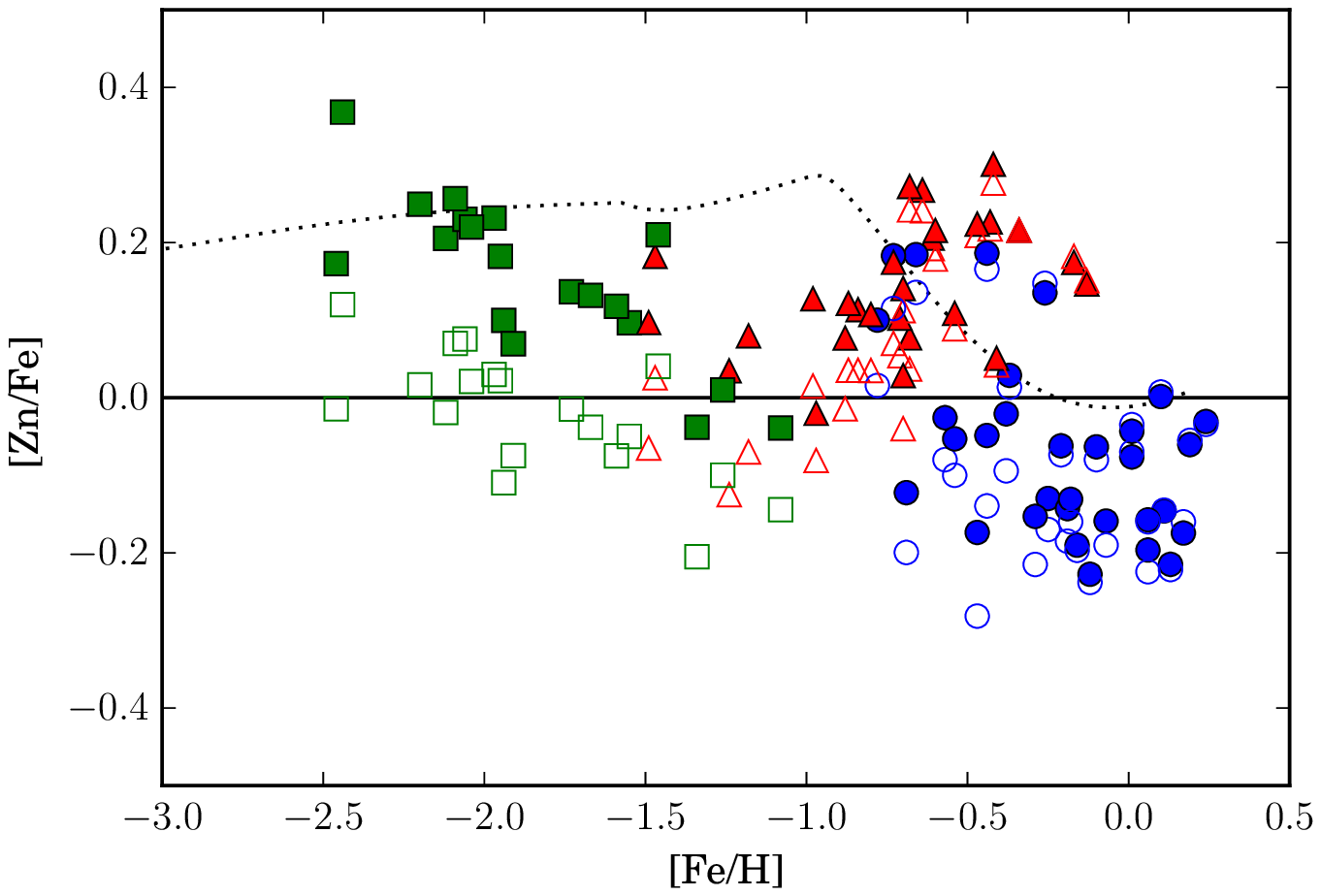}
	\includegraphics[width=80mm]{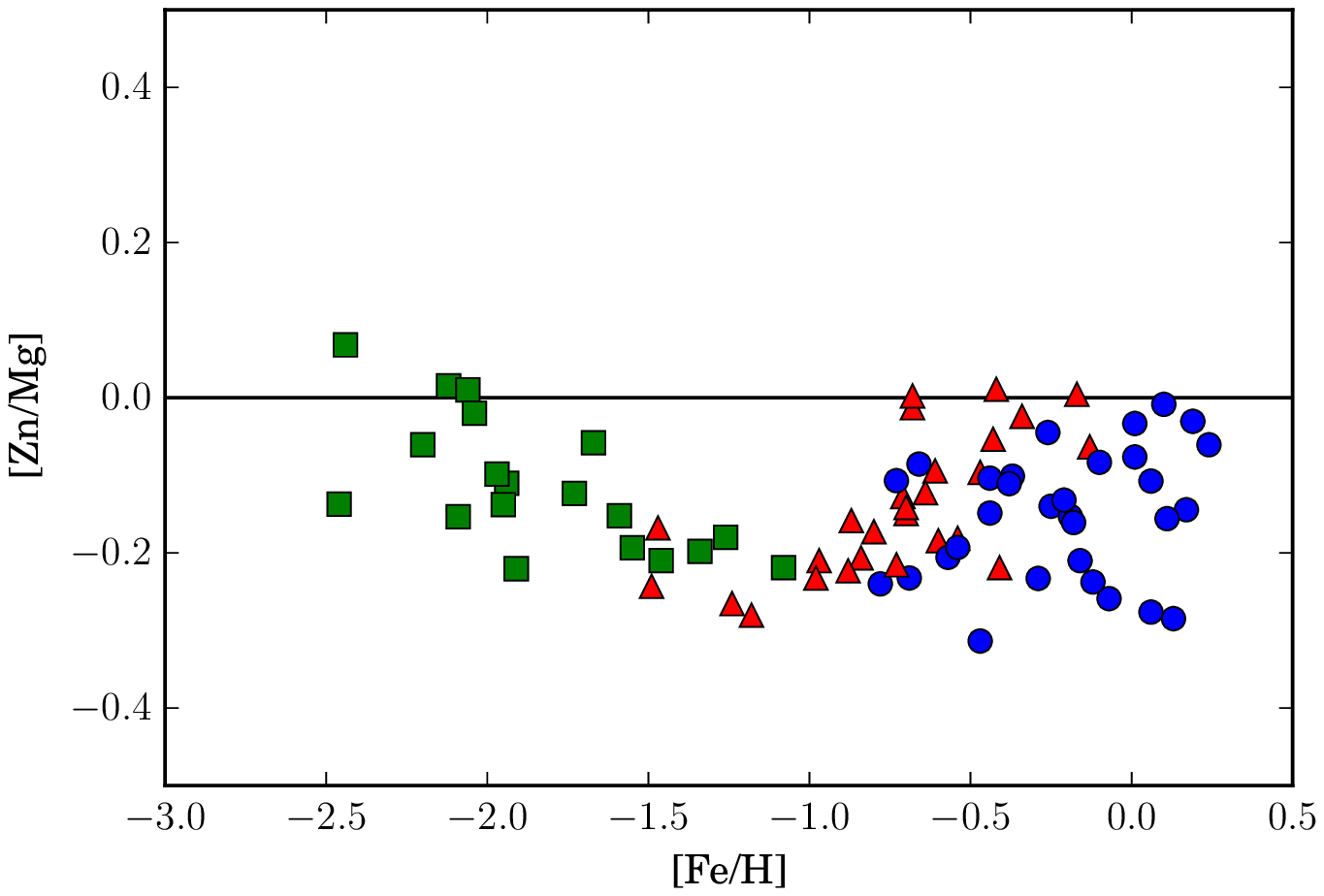}
	\includegraphics[width=80mm]{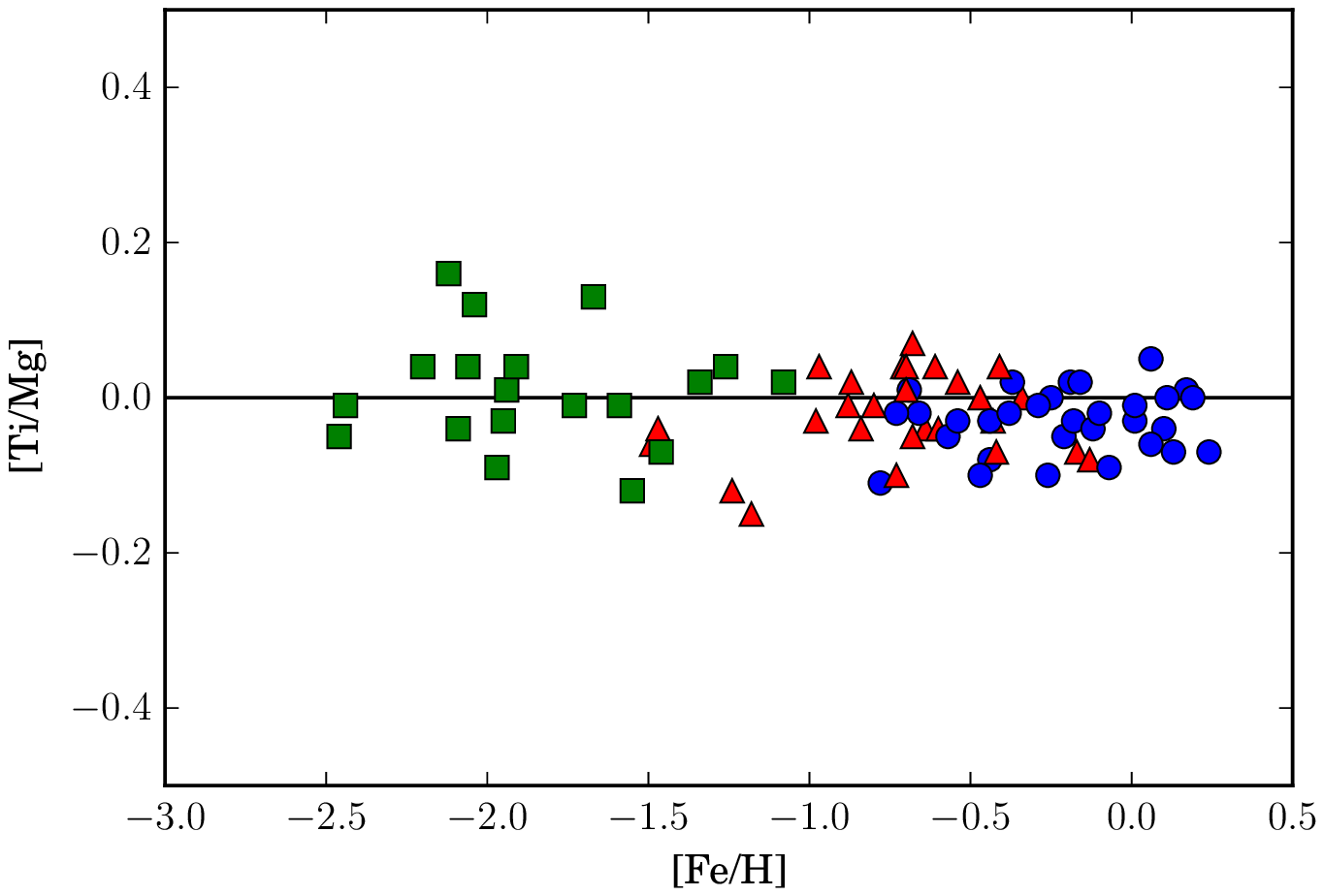}
	\caption{	
		Galactic abundance trends for [Zn/Fe] (top panel), [Zn/Mg] (middle panel), and [Ti/Mg] (bottom panel, for comparison). Halo, thick, and thin disk stars are shown with squares, triangles, and circles, respectively. Non-LTE and LTE abundance ratios are shown with filled and open symbols, respectively. Dotted line shows the Galactic trend predicted by \citet{2020ApJ...900..179K}.}
	\label{trends} 
\end{figure}

We employed Zn\ione\ 4722, 4810, and 6362 \AA\ lines for zinc abundance determination in FGK dwarfs. When fitting the Zn\ione\ 6362 \AA\ line, the Ca\ione\ 6362 \AA\ autoionisation line was taken into account and fitted by an increase in Van Der Waals damping constant. The derived differential LTE and non-LTE abundance ratios are presented in Table~\ref{tab_znh} and Fig.~\ref{trends}. Our zinc abundances are derived from the line-by-line differential approach with respect to the Sun.

Metal-poor stars with $-2.5 <$ [Fe/H] $< -1.2$ show a decrease from  [Zn/Fe]$_{\rm NLTE}$ = 0.3 to 0. 
In the $-1.2 <$ [Fe/H] $< -0.5$ metallicity range, the thick disk stars show an increase in [Zn/Fe]$_{\rm NLTE}$ from 0 to 0.3 dex. Our thin disk stars span $-0.7 <$ [Fe/H] $< 0.2$ and show a decrease in [Zn/Fe] from 0.2 to $-0.2$. The close-to-solar metallicity stars have [Zn/Fe] = $-0.1$, on average.

A number of studies \citep[for example,][]{2003A&A...410..527B,2017A&A...600A..22M} concluded that the observed [Zn/Fe] -- [Fe/H] trend is similar to [$\alpha$/Fe]. 
However, we found that zinc and magnesium abundances do not follow each other (see Fig~\ref{trends}). 
To make sure that the zinc abundance does not behave like the $\alpha$-element abundances, we plotted, for the comparison, the [Zn/Mg]  abundance ratios for our sample stars, using the Mg\ione\  abundances derived by \citet{2016ApJ...833..225Z}, \citet{2019ARep...63..726M}, and \citet[][]{2003A&A...397..275M}. 
We found subsolar [Zn/Mg] in our sample stars. Moreover, [Zn/Mg] trend is not flat: in stars with  $-1.5 <$ [Fe/H] $< -0.7$, an enhancement in zinc is smaller compared to magnesium enhancement, which results in a dip in [Zn/Mg] -- [Fe/H] trend.

We note that throughout all the investigated metallicity range, the [Zn/Fe] spread in  stars with close [Fe/H] is larger compared to those for [Mg/Fe] and [Ti/Fe] in this stellar sample. 
For illustrative purposes, we present [Ti/Mg] -- [Fe/H] trend (Fig.~\ref{trends}) based on our earlier results for the same sample stars \citep{2016ApJ...833..225Z,2019ARep...63..726M,2003A&A...397..275M}. For the stars from \citet{2003A&A...397..275M}, we evaluated [Ti/H] from a differential LTE abundance analysis of the Ti\ii\ 5336 \AA\ line in this study.
[Ti/Mg] -- [Fe/H] trend is flat, with [Ti/Mg] $\simeq$ 0 and the scatter is small throughout all metallicity range.

The derived [Zn/Fe] trend can be understood as follows. At a very metal poor regime, zinc is produced by hypernova explosions \citep[HN,][]{2003Natur.422..871U}. High [Zn/Fe] decreases with increasing [Fe/H] and it reaches [Zn/Fe] = 0 at [Fe/H] = $-1.2$. This behavior argues that the number of HNe is larger at the earlier stages of the Galaxy formation. At [Fe/H] $> -1.2$, zinc is mainly produced in supernova explosions \citep[SN,][]{2006ApJ6531145K}, and their yields depend on metallicity. Zinc production is more efficient in more metal rich SNe. This causes an increase in [Zn/Fe] at $-1.2 <$ [Fe/H] $< -0.8$. For [Fe/H] $> -0.8$, [Zn/Fe] decreases to the solar and subsolar values due to iron production in SN~Ia.
	
We compared the derived [Zn/Fe] trend with predictions of \citet{2020ApJ...900..179K}. Their chemical evolution model reproduces a decrease in [Zn/Fe] for [Fe/H] $> -0.8$. For the lower metallicities, the observed dip at [Fe/H] $\simeq -1.2$ and an uprising trend with decreasing [Fe/H] are not reproduced. It is worth noting that the fraction of HNe is set in the model as a parameter and it was assumed constant. The observed trend at [Fe/H] $< -1$ potentially could be reproduced by applying the metallicity dependent HN fraction. Our observed trend can be helpful for testing and improving the Galactic chemical evolution models.

\begin{table*}
	\caption{Stellar atmosphere parameters and zinc abundance ratios in the sample of dwarfs}
	\label{tab_znh}
	\setlength{\tabcolsep}{0.97mm}
\renewcommand{\arraystretch}{0.77} 
	\begin{tabular}{rrrrrlrrrc}
		\hline
		HD, BD & \teff, & log g & [Fe/H] & \vt & pop. type & $\rm [Zn/H]_{LTE}$ & $\rm [Zn/H]_{NLTE}$& N$_{\rm l}$  \\
               & K &           &         & \kms &          &                    &                    &     &       \\
		\hline
\multicolumn{9}{l}{Atmospheric parameters from \citet{2019ARep...63..726M}:} & \\
        3795 & 5475 & 3.85 & --0.61 & 1.0 & Thick disk & --0.42 $\pm$  0.05 & --0.40 $\pm$  0.04 &   3  & \\ 
10519 & 5740 & 4.01 & --0.64 & 1.1 & Thick disk & --0.40 $\pm$  0.08 & --0.37 $\pm$  0.04 &   3  & \\ 
18757 & 5650 & 4.30 & --0.34 & 1.0 & Thick disk & --0.12 $\pm$  0.06 & --0.12 $\pm$  0.06 &   3  & \\ 
32923 & 5710 & 4.03 & --0.26 & 1.2 & Thin disk  & --0.11 $\pm$  0.03 & --0.12 $\pm$  0.04 &   3  & \\ 
40397 & 5550 & 4.39 & --0.17 & 1.0 & Thick disk &  0.01 $\pm$  0.05 &  0.00 $\pm$  0.06 &   3  & \\ 
55575 & 5960 & 4.29 & --0.37 & 1.2 & Thin disk  & --0.36 $\pm$  0.01 & --0.34 $\pm$  0.02 &   3  & \\ 
64606 & 5280 & 4.63 & --0.68 & 1.0 & Thick disk & --0.64 $\pm$  0.07 & --0.60 $\pm$  0.04 &   3  & \\ 
65583 & 5315 & 4.50 & --0.68 & 0.8 & Thick disk & --0.44 $\pm$  0.13 & --0.41 $\pm$  0.10 &   3  & \\ 
68017 & 5615 & 4.41 & --0.43 & 0.9 & Thick disk & --0.21 $\pm$  0.03 & --0.20 $\pm$  0.03 &   3  & \\ 
69611 & 5940 & 4.17 & --0.60 & 1.2 & Thick disk & --0.42 $\pm$  0.05 & --0.38 $\pm$  0.02 &   3  & \\ 
102158 & 5800 & 4.24 & --0.47 & 1.1 & Thick disk & --0.26 $\pm$  0.06 & --0.25 $\pm$  0.04 &   3  & \\ 
112758 & 5260 & 4.54 & --0.44 & 0.7 & Thin disk  & --0.27 $\pm$  0.06 & --0.25 $\pm$  0.05 &   3  & \\ 
114762 & 5930 & 4.18 & --0.71 & 1.2 & Thick disk & --0.66 $\pm$  0.03 & --0.61 $\pm$  0.03 &   3  & \\ 
132142 & 5100 & 4.47 & --0.42 & 0.7 & Thick disk & --0.14 $\pm$  0.09 & --0.12 $\pm$  0.08 &   3  & \\ 
135204 & 5420 & 4.44 & --0.13 & 0.9 & Thick disk &  0.02 $\pm$  0.07 &  0.02 $\pm$  0.08 &   3  & \\ 
144579 & 5250 & 4.49 & --0.66 & 0.8 & Thin disk  & --0.52 $\pm$  0.04 & --0.48 $\pm$  0.03 &   2  & \\ 
184499 & 5745 & 4.07 & --0.54 & 1.2 & Thick disk & --0.45 $\pm$  0.09 & --0.43 $\pm$  0.06 &   3  & \\ 
201891 & 5900 & 4.29 & --0.97 & 1.2 & Thick disk & --1.05 $\pm$  0.06 & --0.99 $\pm$  0.03 &   3  & \\ 
221830 & 5770 & 4.14 & --0.41 & 1.2 & Thick disk & --0.37 $\pm$  0.05 & --0.36 $\pm$  0.04 &   3  & \\ 
222794 & 5600 & 3.90 & --0.70 & 1.2 & Thick disk & --0.59 $\pm$  0.09 & --0.56 $\pm$  0.05 &   3  & \\ 
\multicolumn{9}{l}{Atmospheric parameters from \citet{lick}:}  & \\
19373 & 6045 & 4.24 &  0.10 & 1.2 & Thin disk &  0.11 $\pm$  0.06 &  0.10 $\pm$  0.06 &   2  & \\ 
22484 & 6000 & 4.07 &  0.01 & 1.1 & Thin disk & --0.02 $\pm$  0.02 & --0.03 $\pm$  0.02 &   2  & \\ 
22879 & 5800 & 4.29 & --0.84 & 1.0 & Thick disk & --0.80 $\pm$  0.02 & --0.73 $\pm$  0.03 &   2  & \\ 
30562 & 5900 & 4.08 &  0.17 & 1.3 & Thin disk &  0.01 $\pm$  0.00 & --0.00 $\pm$  0.00 &   2  & \\ 
34411 & 5850 & 4.23 &  0.01 & 1.2 & Thin disk & --0.06 $\pm$  0.01 & --0.07 $\pm$  0.01 &   2  & \\ 
49933 & 6600 & 4.15 & --0.47 & 1.7 & Thin disk & --0.75 $\pm$  0.08 & --0.64 $\pm$  0.07 &   2  & \\ 
59374 & 5850 & 4.38 & --0.88 & 1.2 & Thick disk & --0.89 $\pm$  0.04 & --0.80 $\pm$  0.02 &   2  & \\ 
59984 & 5930 & 4.02 & --0.69 & 1.4 & Thin disk & --0.89 $\pm$  0.03 & --0.81 $\pm$  0.03 &   2  & \\ 
64090 & 5400 & 4.70 & --1.73 & 0.7 & Halo & --1.74 $\pm$  0.02 & --1.59 $\pm$  0.00 &   2  & \\ 
69897 & 6240 & 4.24 & --0.25 & 1.4 & Thin disk & --0.42 $\pm$  0.01 & --0.38 $\pm$  0.01 &   2  & \\ 
84937 & 6350 & 4.09 & --2.12 & 1.7 & Halo & --2.14 $\pm$  0.04 & --1.91 $\pm$  0.06 &   2  & \\ 
94028 & 5970 & 4.33 & --1.47 & 1.3 & Thick disk & --1.44 $\pm$  0.04 & --1.29 $\pm$  0.05 &   2  & \\ 
102870 & 6170 & 4.14 &  0.11 & 1.5 & Thin disk & --0.03 $\pm$  0.01 & --0.04 $\pm$  0.01 &   2  & \\ 
103095 & 5130 & 4.66 & --1.26 & 0.9 & Halo & --1.36 $\pm$  0.01 & --1.25 $\pm$  0.03 &   2  & \\ 
105755 & 5800 & 4.05 & --0.73 & 1.2 & Thin disk & --0.61 $\pm$  0.01 & --0.55 $\pm$  0.00 &   2  & \\ 
114710 & 6090 & 4.47 &  0.06 & 1.1 & Thin disk & --0.10 $\pm$  0.00 & --0.10 $\pm$  0.00 &   2  & \\ 
134169 & 5890 & 4.02 & --0.78 & 1.2 & Thin disk & --0.76 $\pm$  0.04 & --0.68 $\pm$  0.04 &   2  & \\ 
140283 & 5780 & 3.70 & --2.46 & 1.6 & Halo & --2.47 $\pm$  0.04 & --2.29 $\pm$  0.02 &   2  & \\ 
142091 & 4810 & 3.12 & --0.07 & 1.2 & Thin disk & --0.26 $\pm$  0.10 & --0.23 $\pm$  0.10 &   1  & \\ 
+66 0268 & 5300 & 4.72 & --2.06 & 0.6 & Halo & --1.98 $\pm$  0.04 & --1.83 $\pm$  0.05 &   2  & \\ 
24289 & 5980 & 3.71 & --1.94 & 1.1 & Halo & --2.05 $\pm$  0.03 & --1.84 $\pm$  0.01 &   2  & \\ 
30743 & 6450 & 4.20 & --0.44 & 1.8 & Thin disk & --0.58 $\pm$  0.03 & --0.49 $\pm$  0.02 &   2  & \\ 
43318 & 6250 & 3.92 & --0.19 & 1.7 & Thin disk & --0.37 $\pm$  0.01 & --0.33 $\pm$  0.01 &   2  & \\ 
45067 & 5960 & 3.94 & --0.16 & 1.5 & Thin disk & --0.36 $\pm$  0.12 & --0.35 $\pm$  0.12 &   2  & \\ 
45205 & 5790 & 4.08 & --0.87 & 1.1 & Thick disk & --0.83 $\pm$  0.02 & --0.75 $\pm$  0.03 &   2  & \\ 
52711 & 5900 & 4.33 & --0.21 & 1.2 & Thin disk & --0.28 $\pm$  0.05 & --0.27 $\pm$  0.05 &   2  & \\ 
58855 & 6410 & 4.32 & --0.29 & 1.6 & Thin disk & --0.50 $\pm$  0.01 & --0.44 $\pm$  0.00 &   2  & \\ 
62301 & 5840 & 4.09 & --0.70 & 1.3 & Thick disk & --0.74 $\pm$  0.00 & --0.67 $\pm$  0.00 &   2  & \\ 
74000 & 6225 & 4.13 & --1.97 & 1.3 & Halo & --1.94 $\pm$  0.00 & --1.74 $\pm$  0.02 &   2  & \\ 
76932 & 5870 & 4.10 & --0.98 & 1.3 & Thick disk & --0.96 $\pm$  0.01 & --0.85 $\pm$  0.00 &   2  & \\ 
82943 & 5970 & 4.37 &  0.19 & 1.2 & Thin disk &  0.14 $\pm$  0.01 &  0.13 $\pm$  0.01 &   2  & \\ 
89744 & 6280 & 3.97 &  0.13 & 1.7 & Thin disk & --0.09 $\pm$  0.08 & --0.08 $\pm$  0.08 &   2  & \\ 
90839 & 6195 & 4.38 & --0.18 & 1.4 & Thin disk & --0.34 $\pm$  0.00 & --0.31 $\pm$  0.00 &   2  & \\ 
92855 & 6020 & 4.36 & --0.12 & 1.3 & Thin disk & --0.36 $\pm$  0.06 & --0.35 $\pm$  0.06 &   2  & \\ 
99984 & 6190 & 3.72 & --0.38 & 1.8 & Thin disk & --0.47 $\pm$  0.04 & --0.40 $\pm$  0.03 &   2  & \\ 
100563 & 6460 & 4.32 &  0.06 & 1.6 & Thin disk & --0.16 $\pm$  0.10 & --0.14 $\pm$  0.10 &   2  & \\ 
106516 & 6300 & 4.44 & --0.73 & 1.5 & Thick disk & --0.66 $\pm$  0.00 & --0.56 $\pm$  0.01 &   2  & \\ 
108177 & 6100 & 4.22 & --1.67 & 1.1 & Halo & --1.71 $\pm$  0.06 & --1.54 $\pm$  0.04 &   2  & \\ 
110897 & 5920 & 4.41 & --0.57 & 1.2 & Thin disk & --0.65 $\pm$  0.01 & --0.60 $\pm$  0.02 &   2  & \\ 
115617 & 5490 & 4.40 & --0.10 & 1.1 & Thin disk & --0.18 $\pm$  0.01 & --0.16 $\pm$  0.02 &   2  & \\ 
134088 & 5730 & 4.46 & --0.80 & 1.1 & Thick disk & --0.76 $\pm$  0.01 & --0.69 $\pm$  0.02 &   2  & \\ 
138776 & 5650 & 4.30 &  0.24 & 1.3 & Thin disk &  0.21 $\pm$  0.10 &  0.21 $\pm$  0.10 &   2  & \\ 
142373 & 5830 & 3.96 & --0.54 & 1.4 & Thin disk & --0.64 $\pm$  0.10 & --0.59 $\pm$  0.10 &   1  & \\ 
--4 3208 & 6390 & 4.08 & --2.20 & 1.4 & Halo & --2.18 $\pm$  0.05 & --1.95 $\pm$  0.06 &   2  & \\ 
+7 4841 & 6130 & 4.15 & --1.46 & 1.3 & Halo & --1.42 $\pm$  0.03 & --1.25 $\pm$  0.01 &   2  & \\ 
+9 0352 & 6150 & 4.25 & --2.09 & 1.3 & Halo & --2.02 $\pm$  0.00 & --1.83 $\pm$  0.01 &   2  & \\ 
+24 1676 & 6210 & 3.90 & --2.44 & 1.5 & Halo & --2.32 $\pm$  0.10 & --2.07 $\pm$  0.10 &   1  & \\ 
+29 2091 & 5860 & 4.67 & --1.91 & 0.8 & Halo & --1.98 $\pm$  0.02 & --1.84 $\pm$  0.00 &   2  & \\ 
+37 1458 & 5500 & 3.70 & --1.95 & 1.0 & Halo & --1.93 $\pm$  0.06 & --1.77 $\pm$  0.07 &   2  & \\ 
090--03 & 6007 & 3.90 & --2.04 & 1.3 & Halo & --2.02 $\pm$  0.04 & --1.82 $\pm$  0.06 &   2  & \\ 
\multicolumn{9}{l}{Atmospheric parameters from \citet{2003A&A...397..275M}:}  & [Ti/H]$^1$ \\
29907 & 5500 & 4.64 & --1.55 & 0.6 & Halo & --1.60 $\pm$  0.01 & --1.45 $\pm$  0.00 &   2  & --1.38 \\
31128 & 5980 & 4.49 & --1.49 & 1.2 & Thick disk & --1.55 $\pm$  0.01 & --1.39 $\pm$  0.02 &   2  & --1.21 \\
59392 & 6010 & 4.02 & --1.59 & 1.4 & Halo & --1.66 $\pm$  0.01 & --1.47 $\pm$  0.02 &   2  & --1.33 \\
97320 & 6110 & 4.27 & --1.18 & 1.4 & Thick disk & --1.25 $\pm$  0.00 & --1.10 $\pm$  0.02 &   2  & --0.97 \\
193901 & 5780 & 4.46 & --1.08 & 0.9 & Halo & --1.22 $\pm$  0.04 & --1.12 $\pm$  0.05 &   2  & --0.88 \\
298986 & 6130 & 4.30 & --1.34 & 1.4 & Halo & --1.54 $\pm$  0.01 & --1.38 $\pm$  0.02 &   2  & --1.16 \\
102200 & 6115 & 4.20 & --1.24 & 1.4 & Thick disk & --1.36 $\pm$  0.01 & --1.21 $\pm$  0.01 &   2  & --1.06 \\ 
		\hline
\multicolumn{10}{l}{$^1$ - determined in this study	} \\
	\end{tabular}
\end{table*}

\section{A grid of the non-LTE abundance corrections}\label{grid}

We calculated the non-LTE abundance corrections for cool and hot stars using \textsc {MARCS}  and \textsc {LLmodels}  model atmosphere grids, respectively.
For cool stars, our grid contains non-LTE abundance corrections for the seven lines of Zn\ione\ and the two resonance lines of  Zn\ii. Our stellar parameter range covers models in a wide metallicity range ($-4 \leq$ [Fe/H] $\leq$ 0.5 with a step of 0.5~dex) for dwarfs (5000~K $\leq$ \teff\ $\leq$ 6500~K, 3.0 $\leq$ log~g $\leq$ 5.0, \vt = 1~\kms), subgiants (5000~K $\leq$ \teff\ $\leq$ 5500~K, 2.0 $\leq$ log~g $\leq$ 2.5, \vt = 2~\kms), and giants (4000~K $\leq$ \teff\ $\leq$ 5000~K, 0.5 $\leq$ log~g $\leq$ 2.5, \vt = 2~\kms).
In the calculations of the non-LTE corrections, we adopted [Zn/Fe] = 0.
To estimate an impact of the adopted [Zn/Fe] on the non-LTE corrections, we performed non-LTE calculations for model atmosphere with \teff /log~g/ [Fe/H] = 5500 /4.0 /$-2$ with [Zn/Fe] = 0.2. In this case, we found $\Delta_{\rm NLTE}$(4810) = 0.095, which is similar to our correction $\Delta_{\rm NLTE}$(4810)  = 0.099 found with [Zn/Fe] = 0. 

For hot stars, our grid of non-LTE corrections spans in the following range of stellar parameters: 6600~K $\leq$ \teff\ $\leq$ 11000~K with a step of 200~K for \teff\ $\leq$ 10000~K and 250~K for higher \teff s, 3.0 $\leq$ log~g $\leq$ 5.0 with a step of 0.2~dex, $-0.5 \leq$ [Fe/H] $\leq$ 0.5 with a step of 0.5~dex, and \vt = 2~\kms. The calculations were performed with [Zn/Fe] = 0.
 
Although  we do not analyse Zn\ii\ lines in our sample hot stars, in the literature, there are studies based on Zn\ii\ UV lines \citep[][]{1988ApJ...325..776S,1994A&A...291..521S}. Therefore, we provide the non-LTE corrections for these lines. In A type stars, Zn\ii\ dominates, and, for the resonance Zn\ii\ lines, non-LTE corrections are small. For example, for model atmosphere with \teff\ = 11000~K, log~g = 3, [Fe/H] = 0.0, $\Delta_{\rm NLTE}$(2025, 2062) = 0.06~dex, and the corrections are smaller for model atmospheres with lower \teff\ and larger log~g. Thus, the results of \citet{1988ApJ...325..776S} and \citet{1994A&A...291..521S} are weakly affected by the non-LTE effects.

\begin{figure}
	\includegraphics[width=80mm]{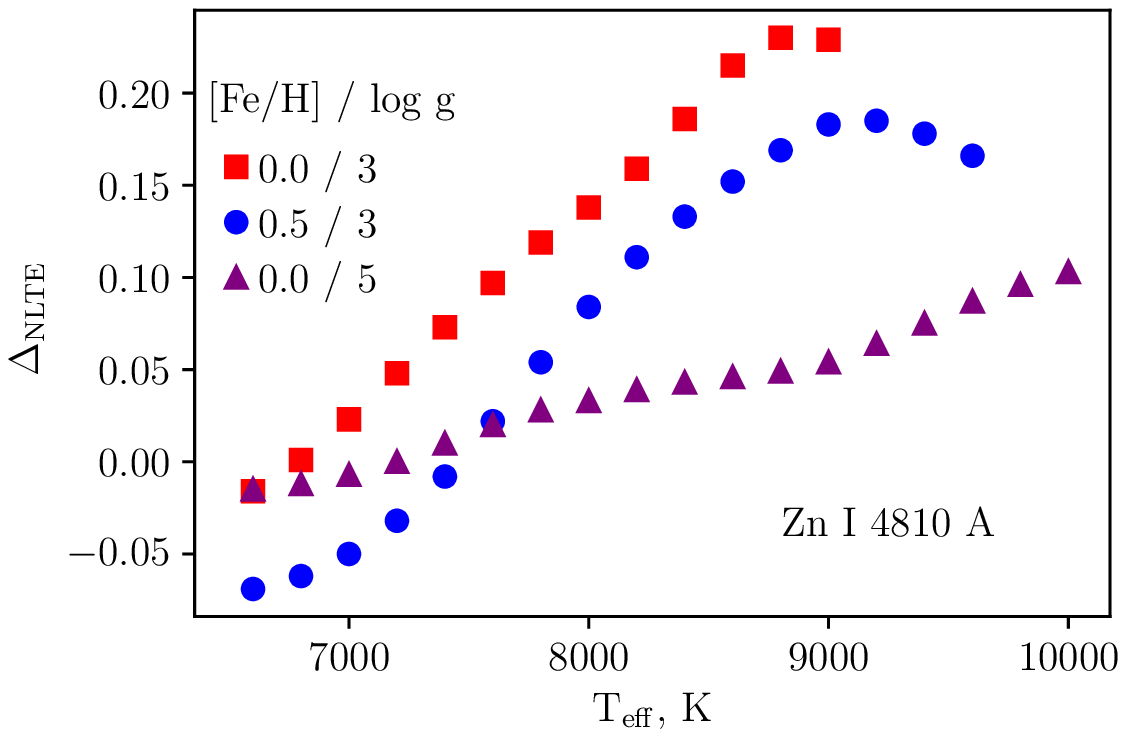}
	\includegraphics[width=80mm]{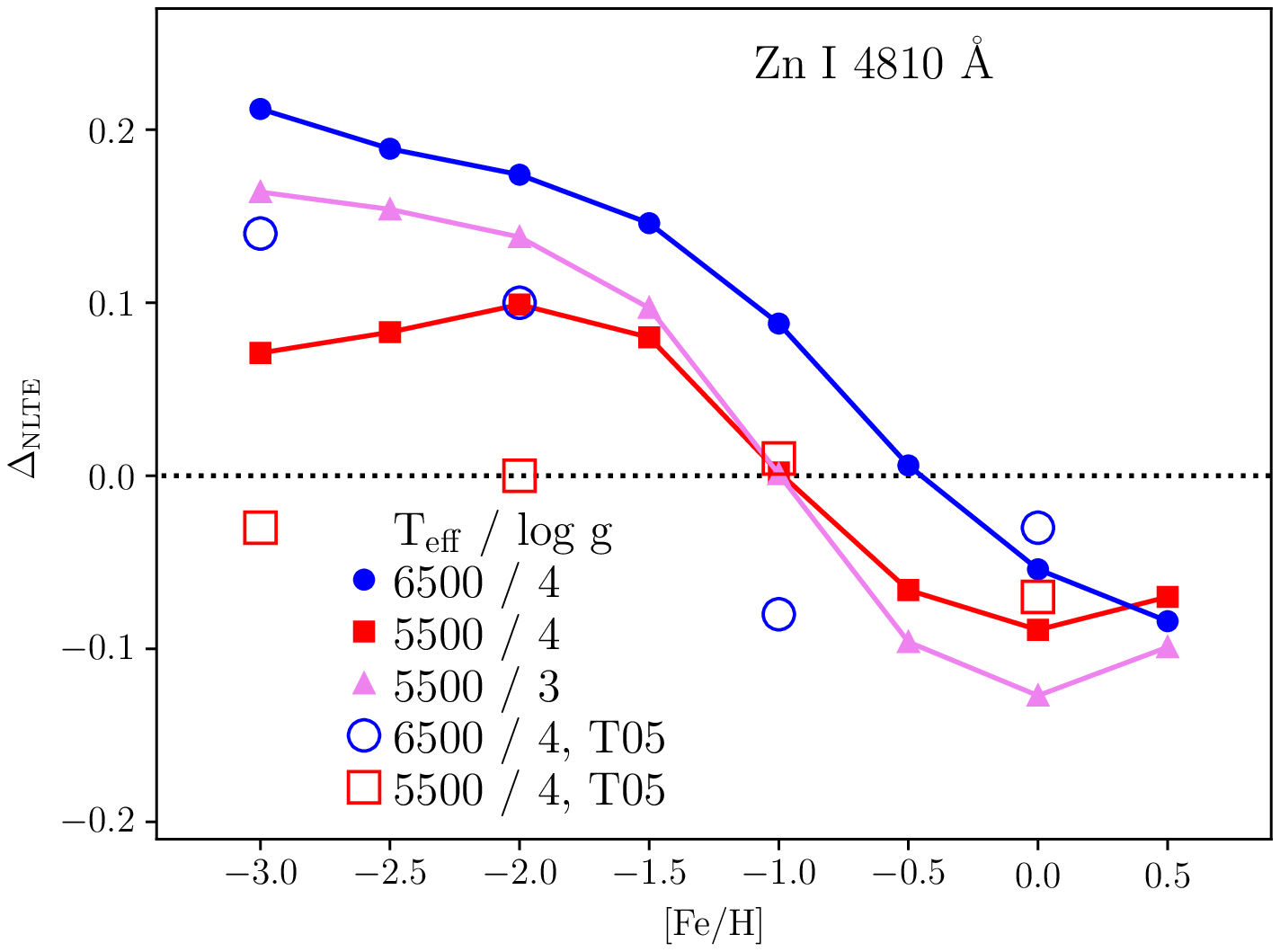}
	\includegraphics[width=80mm]{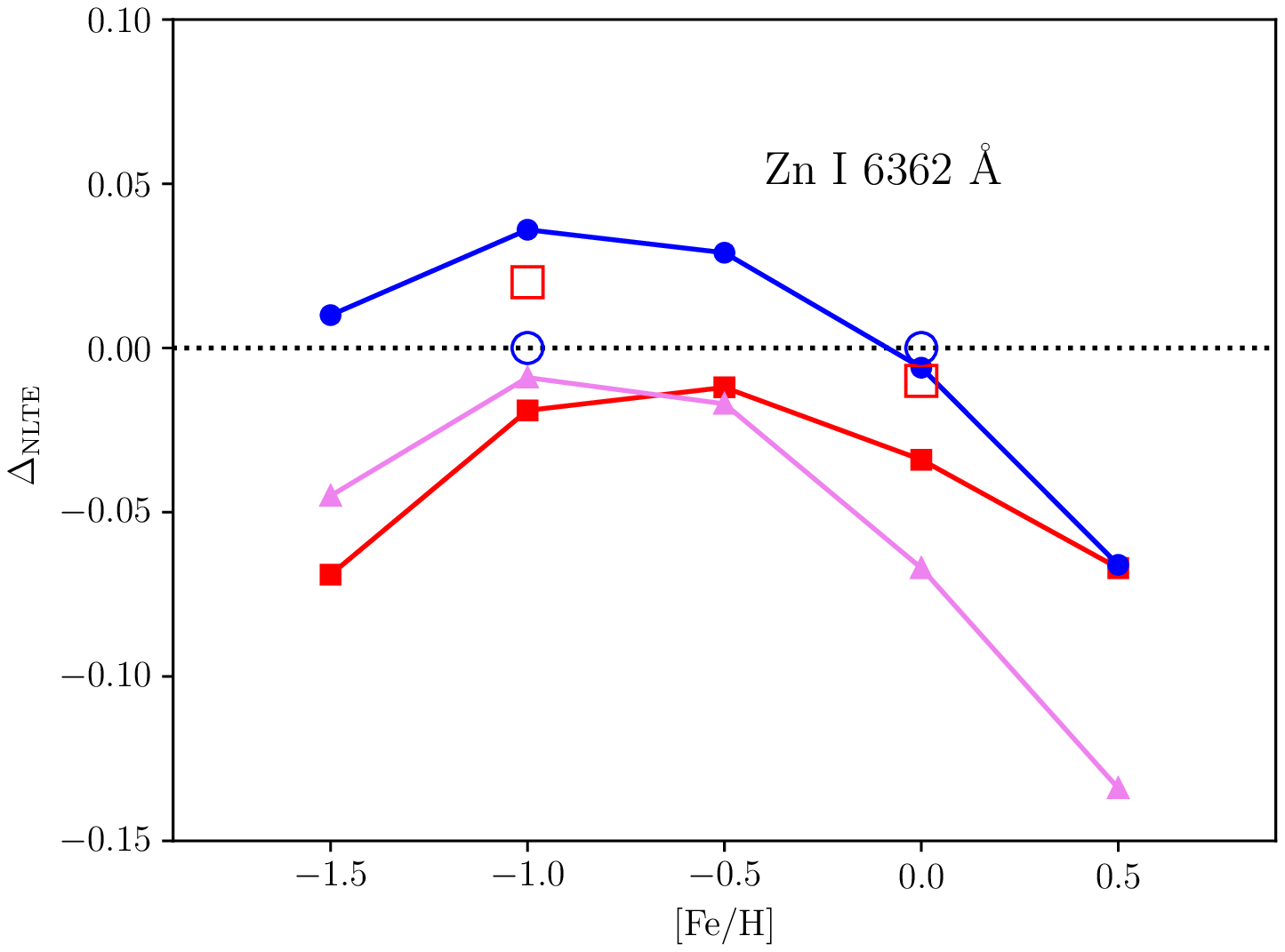}
	\caption{Non-LTE abundance corrections for the selected stellar parameters in a grid of \textsc {LLmodels} (top panel) and \textsc {MARCS} (middle and bottom panels) model atmospheres. The spectral lines are indicated. For comparison, we present non-LTE corrections from T05 (open symbols).}
\label{gridsfig}
\end{figure}

Fig.~\ref{gridsfig} shows the behaviour of the non-LTE abundance corrections for different spectral lines and atmospheric parameters. The correction for Zn\ione\ 4810 \AA\ line strongly depends on the line strength (or metallicity) and it changes its sign at [Fe/H]  $\simeq -1$. Neglecting non-LTE effects when studying stars in a wide metallicity range distorts the Galactic [Zn/Fe] trend (see, for example, open symbols in Fig.~\ref{trends}, top panel).

For model atmospheres with \teff = 5500 K and 6500 K, log~g = 4, and [Fe/H] = 0, $-1$, $-2$, $-3$, we compared the derived non-LTE corrections for Zn\ione\ 4810 and 6362 \AA\ lines with results of T05. For solar metallicity, both studies provide similar results, and the difference in non-LTE corrections is smaller than 0.02~dex (Fig.~\ref{gridsfig}). In metal-poor model atmospheres, there is no agreement between the two studies. For the Zn\ione\ 4810 \AA\ line, our corrections are systematically 0.1~dex larger compared to those of T05. Using different methods, namely, codes, model atoms, and atomic data results in different abundance corrections for metal-poor stars.

\begin{table}
	\label{grid_tab}
	\caption{Non-LTE abundance corrections and equivalent widths for zinc lines as a function of \teff, log~g, and [Fe/H] in MARCS and LL model atmosphere grids.}
\renewcommand{\arraystretch}{0.8} 
	\begin{tabular}{rrrrrrrrrrr}
		\hline
\multicolumn{7}{l}{$\lambda$, nm  species  \eexc, eV  log gf } \\
T$_{\rm eff 1}$, K & log~g$_1$ & log~g$_2$ & ... & log~g$_8$ & log~g$_9$ & log~g$_{10}$ \\
$\rm [Fe/H]_1$ & EW$_1$ & EW$_2$ & ... & EW$_8$ & EW$_9$ & EW$_10$ \\
$\rm [Fe/H]_1$ & $\Delta_1$ & $\Delta_2$ & ... & $\Delta_8$ & $\Delta_9$ & $\Delta_{10}$ \\
\multicolumn{7}{l}{...} \\
		\hline
\multicolumn{7}{l}{213.8573 nm  Zn1  \eexc\ =  0.000  log gf = 0.161} \\
 \teff\ = 5000 & 0.5 & 1.0 & ...  &   4.0  &   4.5  &   5.0\\
  0.5  &     --1 &     --1 & ... &     --1 &     --1 &     --1\\
0.5  & --1.000 & --1.000 & ... & --1.000 & --1.000 & --1.000\\
0.0  &     --1 &     --1 & ... &     --1 &     --1 &     --1\\
0.0  & --1.000 & --1.000 & ... & --1.000 & --1.000 & --1.000\\
--0.5  &     --1 &     --1 & ... &     --1 &     --1 &     --1\\
--0.5  & --1.000 & --1.000 & ... & --1.000 & --1.000 & --1.000\\
--1.0  &     --1 &     --1 & ... &     --1 &     --1 &     --1\\
--1.0  & --1.000 & --1.000 & ... & --1.000 & --1.000 & --1.000\\
--1.5  &     --1 &     --1 & ... &     --1 &     --1 &     --1\\
--1.5  & --1.000 & --1.000 & ... & --1.000 & --1.000 & --1.000\\
--2.0  &     --1 &     --1 & ... &     --1 &     --1 &     --1\\
--2.0  & --1.000 & --1.000 & ... & --1.000 & --1.000 & --1.000\\
--2.5  &     --1 &     --1 & ... &     --1 &     --1 &     --1\\
--2.5  & --1.000 & --1.000 & ... & --1.000 & --1.000 & --1.000\\
--3.0  &     --1 &     --1 & ... &     --1 &     --1 &     --1\\
--3.0  & --1.000 & --1.000 & ... & --1.000 & --1.000 & --1.000\\
--3.5  &     140 &     136 & ... &   207 &   230 &   255 \\
--3.5  &   0.681 &   0.642 & ... & 0.089 & 0.050 & 0.027 \\
--4.0  &      97 &      95 & ... &   123 &   136 &   151 \\
--4.0  &   0.709 &   0.667 & ... & 0.097 & 0.056 & 0.033 \\
		\hline
	\end{tabular}
This table is available in its entirety in a machine-readable form in the online journal. A portion is shown here for guidance regarding its form and content. If EW = $-1$ and $\Delta_{\rm NLTE} = -1$, this means that EW is either $<$ 3~m\AA\ or $>$ 300~m\AA\ in a given model atmosphere.
\end{table}

\section{Conclusions}\label{conclusions}

We investigated the departures from LTE for Zn\ione\--Zn\ii\ in a wide range of stellar parameters with a new model atom of Zn\ione\--Zn\ii\ based on the most up-to-date data on the photoionisation cross-sections, electron-impact excitation rates for Zn\ione, and the Zn\ione\ + H\ione\ and Zn\ii\ + H$^-$ collisions. The latter are treated  with the rate coefficients, which were calculated in this study  for the first time  using the multichannel quantum asymptotic treatment based on the Born-Oppenheimer approach. The non-LTE calculations for Zn\ione\--Zn\ii\ were performed through a range of spectral types from G to late B. 

Using  Zn\ione\ 4722, 4810, and 6362 \AA\ lines, we determined solar zinc abundance log~A$_\odot$ = $-7.40 \pm 0.04$ and $-7.45 \pm 0.02$ in LTE and non-LTE, respectively. For the Sun, non-LTE leads to lower zinc abundance and better agreement between abundances from different lines. Our non-LTE abundance is close to the meteoritic one log~A = $-7.43$ \citep{2021SSRv..217...44L}.

Non-LTE analysis was performed for the first time for the UV Zn\ione\ and Zn\ii\ lines in  two very metal-poor reference stars, HD~84937 and HD~140283.
We found consistent non-LTE abundance from the resonance Zn\ione\ 2138 \AA\ line, the subordinate lines, and the lines of Zn\ii. In both stars, non-LTE leads to 0.17~dex higher average abundance from Zn\ione, while, for Zn\ii\ lines, non-LTE corrections are minor and do not exceed 0.06~dex.	

Using the Zn\ione\ lines in the visible high-resolution spectra, we determined non-LTE abundances for a sample of 80 dwarf stars in the $-2.5 \le$ [Fe/H] $\le$ 0.2 metallicity range. Metal-poor stars with $-2.5 <$ [Fe/H] $< -1.2$ show a decrease from  [Zn/Fe]$_{\rm NLTE}$ = 0.3 to 0. 
In the $-1.2 <$ [Fe/H] $< -0.5$ metallicity range, the thick disk stars show an increase in [Zn/Fe]$_{\rm NLTE}$ from 0 to 0.3 dex. Our thin disk stars span $-0.7 <$ [Fe/H] $< 0.2$ and show a decrease in [Zn/Fe] from 0.2 to $-0.2$. The close-to-solar metallicity stars have [Zn/Fe] = $-0.1$, on average.

We compared the derived [Zn/Fe] trend with predictions of \citet{2020ApJ...900..179K}. 
Their chemical evolution model reproduces a decrease in [Zn/Fe] for [Fe/H] $> -0.5$. For the lower metallicities, the observed dip at [Fe/H] $\simeq -1.2$ and an uprising trend with decreasing [Fe/H]  are not reproduced by the model. 
The derived trend argues for zinc production in the most massive stars and metallicity dependent yields.

For the first time we perform the non-LTE abundance determination of zinc in BAF spectral type stars. For the subordinate Zn\ione\ 4680, 4722, 4810 \AA\ lines, non-LTE leads to either negative, or positive abundance corrections depending on stellar parameters. 
Non-LTE abundances of zinc were derived for the first time for seven reference F to late B-type stars.
HD~32115 and Procyon are deficient in zinc with [Zn/Fe] = $-0.34$ and $-0.11$, respectively. This result is in line with the [Zn/Fe] in  close-to-solar metallicity G dwarfs.

The hotter A and late B-type stars are, in contrast, enhanced in zinc already in LTE and even more in non-LTE. The two of them, Sirius and HD~72660, are known as metallic-line stars and reveal a steep growth of abundances of the heavy elements beyond iron. Their high values of [Zn/H] $\simeq$ 1 lie well in their element abundance patterns.

The remaining three stars, HD~145788, HD~73666, and 21~Peg, reveal similar [Zn/Fe] = 0.22, 0.29, and 0.34, respectively. They have different metallicities because formed in the environments of different chemical composition, but very similar element abundance patterns, which resemble that for the superficially normal stars. However, in these stars, \citet{2020MNRAS.499.3706M} found supersolar abundance ratios for the heavy elements Sr, Zr, Ba, and Nd, and the obtained high abundances of Zn support the abundance trends found by MR20.

To account for deviations from LTE, we provide non-LTE abundance corrections for individual lines of zinc in a wide range of stellar parameters.  

\section*{Acknowledgments}

We thank Klaus Bartschat for providing the data for electron impact excitation of Zn\ione\ in digital form.
The authors are grateful to Klaus Fuhrmann for providing the spectra obtained with the FOCES spectrograph on the 2.2~m telescope of the Calar Alto Observatory. We also thank the referee for the notes and suggestions, which substantially helped to improve the paper.
TS is grateful to Prof. Chiaki Kobayashi for useful comments on the derived Galactic trend. TS acknowledges support from the BASIS Foundation, Project No. 20-1-3-10-1.
SAY gratefully acknowledges support from the Ministry of Education (the Russian Federation), Project No. FSZN-2020-0026. AKB gratefully
acknowledges support from the Basis foundation, Project No. 20-1-1-33-1.
We made use of the StarCAT, NIST, SIMBAD, VALD, and R.~Kurucz databases.

\section{Data availability}

The data used in this article will be shared on request to the corresponding author.

\bibliography{atomic_data,nlte1,stellar_parameter1,mashonkina,zn,ZnH-ref}
\bibliographystyle{mn2e}

\end{document}